\documentclass[aps,onecolumn,showpacs,footinbib]{revtex4}%
\usepackage[latin9]{inputenc}
\usepackage{amssymb}
\usepackage{amsmath}
\usepackage{amscd}
\usepackage{graphicx}
\usepackage{subfigure}
\usepackage{float}
\begin{document}
\draft
\title{Distributed phase-covariant cloning with atomic ensembles via
quantum Zeno dynamics}
\author{Li-Tuo Shen$^{1}$}
\author{Huai-Zhi Wu$^{1}$}
\email{huaizhi.wu@fzu.edu.cn}
\author{Zhen-Biao Yang$^{2}$}

\address{$^{1}$Department of Physics, Fuzhou University, Fuzhou
350002, People's Republic of China\\$^{2}$ Key Laboratory of Quantum
Information, University of Science and Technology of China, CAS,
Hefei 230026, P. R. China}
\date{\today }

\begin{abstract}
We propose an interesting scheme for distributed orbital state
quantum cloning with atomic ensembles based on the quantum Zeno
dynamics. These atomic ensembles which consist of identical
three-level atoms are trapped in distant cavities connected by a
single-mode integrated optical star coupler. These qubits can be
manipulated through appropriate modulation of the coupling constants
between atomic ensemble and classical field, and the cavity decay
can be largely suppressed as the number of atoms in the ensemble
qubits increases. The fidelity of each cloned qubit can be obtained
with analytic result. The present scheme provides a new way to
construct the quantum communication network.
\end{abstract}

\pacs{PACS number: } \maketitle

\vskip 0.5cm

\narrowtext

\section{Introduction}
\label{sec:1} Quantum information processing via quantum Zeno
dynamics \cite{PLA2000275} is a fascinating feature in quantum
communication network, which is different from the quantum Zeno
effect (QZE) \cite{JMP1997756,PRA199041}. The most remarkable
practical application of quantum Zeno dynamics is to suppress
decoherence and dissipation via frequent measurements, which build a
Zeno subspace that the system can evolve in
\cite{PRL200289,JPCS2009012017}. Until now, protocols about
generating entanglement and implementing the quantum logic gate via
quantum Zeno dynamics have been proposed
\cite{PRA2008062332,PRA2003022320,PRA2007052339,EPL201050003,PRA2009062323,PRA2011022322},
and the QZE has been experimentally observed in various systems
\cite{PRT198821,PRL2008180402,PRL2006260402}, such as cavity QED
\cite{PRL2008180402} and Bose-Einstein condensates
\cite{PRL2006260402}. On the other hand, Pellizzari
\cite{PRL19975242} first proposed a scheme to realize the reliable
transfer of quantum information in two distant cavities connected by
an optical fiber in 1997, which provides an essential tool for
long-distant quantum communication schemes in recent years
\cite{APL2009154101,PRA2007012324,PRL2006010503,PRA2009012305,PRA2008014303,PRA2010042327,PRA2007062320,EPL200760001,PRL2004210501}.
Yin and Li \cite{PRA2007012324} generalized the idea to the atomic
ensemble case, where the operation time of quantum computation can
be greatly speeded up, however, the scheme is based on the
coherently dynamical evolution which is more sensitive to the
effects of spontaneous emission of atoms and photon leakage out of
cavity. Unfortunately, there is still no report about distributed
quantum computation via quantum Zeno dynamics with the atomic
ensembles.

Another distinguishing feature of quantum system is quantum state
cloning, which is first proved by Wootters and Zurek
\cite{Nature1982299} as the no-cloning theorem that based on the
linearity of quantum mechanics. However, a lot of research about
approximate quantum state cloning has been proposed
\cite{PRA2000012302,PRA2001012304,PRA2003042306,PRA199858,PRA199654,PRL199779,JOB2005139,Nature2011210,IEEE2004900,PLA2005278}.
Many experimental realizations of quantum state cloning have also
been done
\cite{Science2002296,PRL2002187901,PRL2005040505,PRL2011180404,PRL2007170503,PRL2010073602}.
But the proposal about distributed quantum state cloning is still
not reported.

To avoid the defects that decoherence is a main influence in the
previous system taking one atom as a qubit, and to take full
advantages of the atomic ensembles in reducing the effect of
dissipation, we propose a scheme for distributed quantum state
cloning with atomic ensembles via quantum Zeno dynamics. Compared
with the previous schemes with only one atom as a qubit, the
distinct features of our scheme are as follows: (1) The entanglement
of the atomic ensembles is very robust against cavity decay and the
distributed orbital state quantum cloning can be realized with only
one step without auxiliary qubit. (2) The coupling strength between
the atom and cavity can be collectively enhanced as the number of
the atoms in each cavity increases, but the operation time prolongs
much slowly. Thus, the demand on the rigorous condition of large
coupling strength between single atom and cavity field becomes not
so high in our scheme. (3) The quantum state cloning can be
controlled by adjusting classical laser at different nodes. The
present scheme can also be generalized to other systems, such as
superconducting quantum interference device \cite{Nature200653} and
nitrogen vacancy center \cite{Nature2010249}. This kind of quantum
cloning is important tools for studying a wide variety of task, such
as state estimation \cite{PRL19982598}, quantum cryptography
\cite{PRL1991661} and distributed measurements \cite{PRA2001042308},
and provides a very new way for the quantum communication network in
future.

\section{Model} The quantum Zeno dynamics is governed by the Hamiltonian
$H_K=H+KH_m$, where $H$ is the Hamiltonian of the quantum system to
be investigated and $H_m$ can be considered as an additional
interaction Hamiltonian performing the measurement. $K$ stands for
coupling constant. The subsystem of interest is governed by the
evolution operator
$U(t)=\lim_{K\rightarrow\infty}\exp(iKH_mt)U_K(t)$ when a strong
coupling limit $K\rightarrow\infty$, which has the form
$U(t)=\exp(-it\sum_{n}P_nHP_n)$ and $P_n$ represents the
eigenprojection of the $H_m$ corresponding to the eigenvalue
$\lambda_n$
($H_m=\sum_n\lambda_nP_n$)~\cite{PRL200289,JPCS2009012017}. Thus the
limiting evolution operator $U_K(t)$ $\sim$ $\exp(-iKH_mt)U(t)$ =
$\exp[-i\sum_n(K\lambda_nP_n+P_nHP_n)t]$, so that the effective
Hamiltonian can be written as
$H_{eff}=\sum_n(K\lambda_nP_n+P_nHP_n)$, which is an important
result in the quantum Zeno dynamics.

As shown in Fig. 1 (a), we consider that the atomic ensembles are
trapped in each of the distant cavities connected by a $1\times N$
($N$ $\geq$ $3$) single-mode integrated optical star coupler
\cite{APL1982549,JPNJAP19975136}, where each atomic ensemble
consists of $M$ identical three-level atoms. The optical star
coupler is made up of $N$ identical optical fiber channels and only
one resonant field mode interacts with the cavity mode which
possesses the ability for present quantum information processing.
Under the rotating-wave approximation, the Hamiltonian of the whole
system in the interaction picture reads
\begin{eqnarray}\label{eq.1}
H_{total}&=&H_{laser}+H_{I},\cr
H_{laser}&=&\sum_{x=1}^{N}\sum_{i=1}^{M}\Omega_x^{}|e_{i,x}\rangle\langle
f_{i,x}|+H.c.,\cr\cr
H_{I}&=&\sum_{x=1}^{N}\sum_{i=1}^{M}(g_{x}^{}a_{x}|e_{i,x}\rangle\langle
g_{i,x}|+v_{x}^{}b^+a_{x})+H.c.,
\end{eqnarray}
where $H_{laser}$ plays a role of the Hamiltonian to be investigated
and $H_{I}$ acts as an additional interaction Hamiltonian performing
the measurement in quantum Zeno dynamics. Here, the notation
$|K_{i,x}\rangle$ ($K=e,f,g$) represents atom $i$ with the state
$|K\rangle$ in cavity $x$. The transition $|f_{i,x}\rangle$
$\leftrightarrow$ $|e_{i,x}\rangle$ is resonantly driven by the
classical laser field with the Rabi frequency $\Omega_x$, the atomic
transition $|g_{i,x}\rangle$ $\leftrightarrow$ $|e_{i,x}\rangle$ is
resonantly coupled to the field mode of the $x$th cavity with the
coupling constant $g_x$. $a_x$ ($b$) and $a_x^+$ ($b^+$) are the
annihilation and creation operators for the $x$th cavity field (the
field mode of the optical fiber channel), and $v_x$ is the coupling
strength of the $x$th cavity mode to the field mode of the optical
fiber channel. We assume the parameters $g_{x}=g$, $v_{x}=v$ and
$\Omega_{x^{'}}=\Omega$ ($x^{'}$ = 2, 3,...,\emph{N}) for
simplicity. The interaction between atomic ensemble and cavity field
is collectively enhanced leading to a energy splitting of
$2\sqrt{M}g$, as shown in Fig. 1 (b). The driving laser at the
transition resonant frequency $w$ is detuning from the both dressed
states. In consequence, the atom-cavity system is only perturbed by
the classical field driving if the driving strength is weak enough
comparing with the energy splitting, which is the fundamental
principle of the quantum Zeno dynamics. For convenience, in the
following we take the expression $|A_{f},B_{g},C_{e}\rangle_x$ ($A$,
$B$, $C$ = $0$, $1$, $2$, ...) that denotes the state of atomic
ensemble, where there are $A$ atoms, $B$ atoms and $C$ atoms in the
ground state $|f\rangle$, $|g\rangle$ and $|e\rangle$ of each atom
in the $x$th cavity, respectively. $| 1,M-1,0\rangle _x$ denotes the
symmetric superposition of the states for which only one atom is in
$|f\rangle $ and the other in $|g\rangle $ in the xth atomic
ensemble, i.e., the W state \cite{JOB2005139}.

Assume that the first ensemble is initially in the W state and all
the atoms in other ensembles are initially in $| g\rangle$. The
initial state of the whole system is
\begin{eqnarray}\label{eq.2}
|\Phi(0)\rangle&=&|1_f,M-1_g,0\rangle_{1}\otimes\prod_{\xi=2}^{N}|0_f,M_g,0_e\rangle_{\xi}\cr&&
\otimes\prod_{i=1}^{N}|0\rangle_{i}\otimes|0\rangle_f,
\end{eqnarray}
Thus, the system will evolve in the subspace
\begin{eqnarray}\label{eq.3}
\mathcal {L} &=& \bigg\{ (|\phi_{1}\rangle, |\phi_{2}\rangle,
|\phi_{3}\rangle) , |\phi_{4}\rangle, (|\phi_{5}\rangle,
|\phi_{6}\rangle, |\phi_{7}\rangle) , \cr&& (|\phi_{8}\rangle,
|\phi_{9}\rangle, |\phi_{10}\rangle) ,  ... , (|\phi_{3N-1}\rangle,
|\phi_{3N}\rangle, |\phi_{3N+1}\rangle) \bigg\} \cr &\equiv& \bigg
\{ ( |1_f,M-1_g,0_e\rangle_{1}|0\rangle_1,
|0_f,M-1_g,1_e\rangle_{1}|0\rangle_1,\cr&&
|0_f,M_g,0_e\rangle_{1}|1\rangle_1 ) \otimes\prod_{i=2}^{N}
|0_f,M_g,0_e\rangle_{i} |0\rangle_i \otimes |0\rangle_{f},\cr&&
\prod_{i=1}^{N} |0_f,M_g,0_e\rangle_{i} |0\rangle_i \otimes
|1\rangle_{f}, \cr&& (|0_f,M_g,0_e\rangle_{2}|1\rangle_2,
|0_f,M-1_g,1_e\rangle_{2}|0\rangle_2,\cr&&
|1_f,M-1_g,0_e\rangle_{2}|0\rangle_2 )\otimes \prod_{i=1,i\neq2}^{N}
|0_f,M_g,0_e\rangle_{i} |0\rangle_i \otimes |0\rangle_{f},\cr&&
(|0_f,M_g,0_e\rangle_{3}|1\rangle_3,
|0_f,M-1_g,1_e\rangle_{3}|0\rangle_3,\cr&&
|1_f,M-1_g,0_e\rangle_{3}|0\rangle_3 )
 \otimes
\prod_{i=1,i\neq3}^{N} |0_f,M_g,0_e\rangle_{i} |0\rangle_i \otimes
|0\rangle_{f},\cr&& ..., \cr&& (|0_f,M_g,0_e\rangle_{N}|1\rangle_N,
|0_f,M-1_g,1_e\rangle_{N}|0\rangle_N,\cr&&
|1_f,M-1_g,0_e\rangle_{N}|0\rangle_N) \otimes \prod_{i=1}^{N-1}
|0_f,M_g,0_e\rangle_{i} |0\rangle_i \otimes |0\rangle_{f}
\bigg\}.\cr&&
\end{eqnarray}
Under the condition of quantum Zeno dynamics $\Omega$, $\Omega_1$
$\ll$ $g$, $v$, the whole Hilbert subspace is split into different
invariant Zeno subspaces, the eigenprojection of $H_{I}$ is
$P^{\alpha}_{n}$ = $|\alpha\rangle\langle\alpha|$
($n=1,2,3,...,3N+1$), where $|\alpha\rangle$ is the corresponding
eigenstate of $H_{I}$ in the whole subspace. If the corresponding
eigenvalue is $\lambda_{n}$, then the effective Hamiltonian reads
\cite{JPCS2009012017}
\begin{eqnarray}\label{e.4-e.5}
H_{total}&\simeq&
\sum_{n,\alpha,\beta}(\lambda_{n}P_{n}^{\alpha}+P_{n}^{\alpha}H_{laser}P_{n}^{\beta})\cr&&=\sum_{n,\alpha}\lambda_{n}P_{n}^{\alpha}+H_{e},\\
H_{e}&=&\frac{v}{\sqrt{Nv^{2}+Mg^{2}}}\bigg(\Omega_1^{}|\phi_{1}\rangle\langle\varphi_{2}|\cr&&+\Omega^{}\sum_{k=2}^{N}|\phi_{3k+1}\rangle\langle\varphi_{2}|+H.c.\bigg),
\end{eqnarray}
where
\begin{eqnarray}\label{eq.6}
|\varphi_{2}\rangle =
\frac{v^{}}{\sqrt{Nv^{2}+Mg^{2}}}(|\phi_{2}\rangle-\frac{\sqrt{M}g^{}}{v^{}}|\phi_{4}\rangle+\sum_{k=2}^{N}|\phi_{3k}\rangle),
\end{eqnarray}
after an interaction time $t$, the state of the system becomes
\begin{eqnarray}\label{eq.7}
|\Phi^{}(t_{})\rangle&=&\bigg[\frac{\Omega_{1}^{}}{\Omega^{}}+\frac{(N-1)\Omega^{}}{\Omega_1^{}}\bigg]^{-1}
\bigg\{\bigg[\frac{\Omega_1^{}}{\Omega^{}}\cos(\mu t)\cr&&
+\frac{(N-1)\Omega^{}}{\Omega_1^{}}\bigg]  |\phi_{1}\rangle
+\bigg[\cos(\mu t)-1\bigg]\sum_{k=2}^{N}|\phi_{3k+1}\rangle\cr&&
-i\frac{\sqrt{\Omega_1^{2}+(N-1)\Omega^{2}}}{\Omega^{}}\sin(\mu
t)|\varphi_{2}\rangle\bigg\},
\end{eqnarray}
where $\mu$ =
$\bigg(v^{}\sqrt{\Omega_1^{2}-\Omega^{2}+N\Omega^{2}}\bigg)$ $/$
$\sqrt{Nv^{2}+Mg^{2}}$. If we set $t$ = $(2n+1)$ $\pi$ / $\mu$ ($n$
= $0, 1, 2, ...$) and $\Omega_1^{}$ =$(\sqrt{N}+1)\Omega^{}$, then
the system will become
\begin{eqnarray}\label{eq.8}
|\Phi^{}(t)\rangle&=&
\frac{1}{\sqrt{N}}\bigg(|\phi_{1}^{}\rangle+\sum_{k=2}^{N}|\phi_{3k+1}^{}\rangle\bigg)\cr\cr
&=&\frac{1}{\sqrt{N}}\bigg(|1_f,M-1_g,0_e\rangle_{1}|0_f,M_g,0_e\rangle_{2}\cr&&|0_f,M_g,0_e\rangle_{3}...|0_f,M_g,0_e\rangle_{N}
\cr&&+|0_f,M_g,0_e\rangle_{1}|1_f,M-1_g,0_e\rangle_{2}\cr&&|0_f,M_g,0_e\rangle_{3}...|0_f,M_g,0_e\rangle_{N}
\cr&&+|0_f,M_g,0_e\rangle_{1}|0_f,M_g,0_e\rangle_{2}\cr&&|1_f,M-1_g,0_e\rangle_{3}...|0_f,M_g,0_e\rangle_{N}
\cr&&+\cdot\cdot\cdot
\cr&&+|0_f,M_g,0_e\rangle_{1}|0_f,M_g,0_e\rangle_{2}\cr&&|0_f,M_g,0_e\rangle_{3}...|1_f,M-1_g,0_e\rangle_{N}\bigg)
\cr&&\otimes\prod_{i=1}^{N}|0\rangle_{i}\otimes|0\rangle_f.
\end{eqnarray}

Therefore the coupling strength between the atom and cavity can be
collectively enhanced to $\sqrt{M}g$, but the operation time
prolongs much slowly. Therefore, we obtain the $W$ state of the
atomic ensembles in \emph{N} separate cavities.  It should be noted
that the key step to achieve the orbital state quantum cloning is to
prepare the $W$ state of the atomic ensembles \cite{JOB2005139}. For
this reason, it is necessary to consider the feasibility for
generating the $W$ state of the atomic ensemble qubits. In the above
derivations, we assume the condition of quantum Zeno dynamics
$\Omega_1$, $\Omega$ $\ll$ $g'$, $v$, therefore, we take the
influence of the ratios $\Omega/(\sqrt{M}g)$ and $v/(\sqrt{M}g)$ on
the fidelity of the $W$ state of atomic ensemble into consideration,
as shown in Fig. 2 (a). The result shows that the fidelity decreases
in an oscillating form with the increase in the ratio of $\Omega/g'$
and keeps higher than $90\%$ even when $\Omega$ $=$ $0.1\sqrt{M}g$
and $v$ = $0.5\sqrt{M}g$, hence our scheme can work well in a large
scale of feasible numbers for $v$ and $\Omega$. In Fig. 2 (b), We
plot the fidelity $F$ versus the dimensionless parameters
$\kappa/(\sqrt{M}g)$, $\gamma/(\sqrt{M}g)$ and $\beta/(\sqrt{M}g)$,
where $\gamma$, $\kappa$, $\beta$ are, respectively, the decay rates
for the spontaneous emission of atom in each cavity and photon
leakage out of each cavity and the optical fiber channel. The result
indicates that the dominant factor of reducing the fidelity is the
atomic spontaneous emission, and the fidelity is almost unaffected
by the decay of the cavity even when $\kappa$ $=$ $0.01\sqrt{M}g$.
If we fix the parameter $g$, the ratio $\Omega/(\sqrt{M}g)$,
$\kappa/(\sqrt{M}g)$, $\gamma/(\sqrt{M}g)$ and $\beta/(\sqrt{M}g)$
all become tunable by choosing the suitable parameter $M$, which is
simpler and more feasible under the current experiment condition. We
also consider the main experiment setups to implement the
distributed quantum state cloning. The fiber-based high-finesse
cavity parameters $(g^{'}, \kappa, \gamma)/2\pi$ = $(185, 53, 3)$MHz
have been reported in the recent experiment \cite{Nature2011210}.
When the optical fiber loss factor at the $852nm$ wavelength is
$2.2$ $dB$ $km^{-1}$ \cite{IEEE2004900}, corresponding to the decay
rate $\beta$ = $0.15$MHz. In a general case in the present scheme,
for example $M$ =100, then the real coupling constant $g$ between
each atom and the cavity is only $18.5$MHz, so that $v$ =
$0.5\sqrt{M}g$ = $90$MHz which is not so large in experiment. With
decoherence of the quantum system, the distributed multi-atom $W$
states generation scheme can be obtained with a high fidelity larger
than $97.66\%$ when $N$ = 3, and the interaction time $t$ needed to
complete the operation is $0.147us$. The deviation $|\delta g^{'}|$
= $0.1g^{'}$ only reduces the fidelity by about $10^{-2}$ when $N$ =
$3$, where $g^{'}$ = $\sqrt{M}g$ and the deviation may be induced by
the number of atoms in each cavity or the coupling strength between
single atom and the cavity.

Based on the previously prepared $W$ state, the quantum cloning
scheme is now able to be implemented. Again, we assume that all the
cavities and the optical fiber channel are initially both in the
vacuum state, and only one atom of the first cavity is prepared in
the arbitrary orbital state of the Bloch sphere \cite{PLA2005278}
\begin{eqnarray}\label{eq.9}
|\psi_\rangle=\cos(\frac{\theta}{2})|g_{}\rangle+\sin(\frac{\theta}{2})e^{i\delta}|f_{}\rangle,
\end{eqnarray}
where the angle $\theta$ is a known parameter, but $\delta$ is
unknown to us, and the other atoms are initially in the ground
states $|g\rangle$. We assume that the first ensemble is initially
in the superposition of $|0,M,0\rangle $ and $|1,M-1,0\rangle $ and
all the atoms in other ensembles are initially in $|g\rangle $. Thus
the whole system is initially prepared in the state
\begin{eqnarray}\label{eq.10}
|\Psi^{}(0)\rangle&=&\bigg(\cos(\frac{\theta}{2})|0_f,M_g,0_e\rangle_{1}+\sin(\frac{\theta}{2})e^{i\delta}|1_f,M-1_g,0_e\rangle_{1}
\bigg)\cr&&\otimes\prod_{\xi=2}^{N}|0_f,M_g,0_e\rangle_{\xi}
\otimes\prod_{i=1}^{N}|0\rangle_{i}\otimes|0\rangle_f.
\end{eqnarray}
The state $\prod_{i=1}^{N}$ $|0_f,M_g,0_e\rangle_{i}$
$|0\rangle_{i}\otimes|0\rangle_f$ undergoes no changes because the
corresponding effective Hamiltonian $H^{'}_{e}$ = 0 in the quantum
Zeno dynamics. On the other hand, the state
$|1_f,M-1_g,0_e\rangle_{1}$ $\otimes\prod_{\xi=2}^{N}$
$|0_f,M_g,0_e\rangle_{\xi}$ $\otimes\prod_{i=1}^{N}$
$|0\rangle_{i}\otimes|0\rangle_f$ will evolve in the subspace
$\mathcal {L}$.

The effective reduced density operators $\rho_{eff}^{1}$ of qubit
$1$ and $\rho_{eff}^{2}$ of qubit $2$ are
\begin{eqnarray}\label{eq.11-eq.12}
\rho_{eff}^{1}&=&\cos^4(\frac{\theta}{2})+|A(t)|^2\sin^4(\frac{\theta}{2})+\cos^2(\frac{\theta}{2})\sin^2(\frac{\theta}{2})
\cr&&\bigg[
(N-1)(|B(t)|^2+|C(t)|^2)\cr&&+|D(t)|^2+A(t)+A^{\dag}(t)\bigg],\\
\rho_{eff}^{2}&=&\cos^4(\frac{\theta}{2})+|B(t)|^2\sin^4(\frac{\theta}{2})+\cos^2(\frac{\theta}{2})\sin^2(\frac{\theta}{2})
\cr&&\bigg[(N-2)(|B(t)|^2+|C(t)|^2)+|A(t)|^2\cr&&+|C(t)|^2+|D(t)|^2+B(t)+B^{\dag}(t)\bigg],
\end{eqnarray}
where
\begin{eqnarray}\label{eq.13}
A(t)&=&\bigg[\frac{\Omega_{1}^{}}{\Omega^{}}+\frac{(N-1)\Omega^{}}{\Omega_1^{}}\bigg]^{-1}
\bigg[\frac{\Omega_1^{}}{\Omega^{}}\cos(\mu t)
+\frac{(N-1)\Omega^{}}{\Omega_1^{}}\bigg],\cr
B(t)&=&\bigg[\frac{\Omega_{1}^{}}{\Omega^{}}+\frac{(N-1)\Omega^{}}{\Omega_1^{}}\bigg]^{-1}
\bigg[\cos(\mu t)-1\bigg],\cr C(t)&=&-i\sin(\mu
t)\frac{\sqrt{\Omega_1^{2}+(N-1)\Omega^{2}}}{\Omega^{}}\frac{v^{}}{\sqrt{Nv^{2}+Mg^{2}}},\cr
D(t)&=&i\sin(\mu
t)\frac{\sqrt{\Omega_1^{2}+(N-1)\Omega^{2}}}{\Omega^{}}\frac{\sqrt{M}g}{\sqrt{Nv^{2}+Mg^{2}}},
\end{eqnarray}
and other effective reduced density operator $\rho_{eff}^{j}$ for
qubit $j$ ($j\geq3$) is exactly the same with that of qubit $2$,
this is because the encoding in qubit $2$ is the exactly symmetry to
that in qubit $j$, therefore we just focus on the qubit $1$ that is
to be cloned and qubit $2$ that has been cloned in the following
discussions. Set $t^{'}$ = $\pi\sqrt{Nv^{2}+g^{2}}$ $/$
$\bigg(v^{}\sqrt{\Omega_1^{2}-\Omega^{2}+N\Omega^{2}}\bigg)$ and
$\Omega_1^{}$ =$(\sqrt{N}+1)\Omega^{}$, the whole system becomes
\begin{eqnarray}\label{eq.14}
|\Psi^{}(t^{'})\rangle&=&
[\cos\frac{\theta}{2}\prod_{i=1}^{N}|0_f,M_g,0_e\rangle_{i}+\sin\frac{\theta}{2}e^{i\delta}\frac{1}{\sqrt{N}}\cr&&
\bigg(|1_f,M-1_g,0_e\rangle_{1}|0_f,M_g,0_e\rangle_{2}\cr&&|0_f,M_g,0_e\rangle_{3}...|0_f,M_g,0_e\rangle_{N}
\cr&&+|0_f,M_g,0_e\rangle_{1}|1_f,M-1_g,0_e\rangle_{2}\cr&&|0_f,M_g,0_e\rangle_{3}...|0_f,M_g,0_e\rangle_{N}
\cr&&+|0_f,M_g,0_e\rangle_{1}|0_f,M_g,0_e\rangle_{2}\cr&&|1_f,M-1_g,0_e\rangle_{3}...|0_f,M_g,0_e\rangle_{N}
\cr&&+\cdot\cdot\cdot
\cr&&+|0_f,M_g,0_e\rangle_{1}|0_f,M_g,0_e\rangle_{2}\cr&&|0_f,M_g,0_e\rangle_{3}...|1_f,M-1_g,0_e\rangle_{N}\bigg)
\cr&&\otimes\prod_{i=1}^{N}|0\rangle_{i}\otimes|0\rangle_f,
\end{eqnarray}
the corresponding fidelity of each cloned qubit reads
\begin{eqnarray}\label{eq.15}
F&=&\cos^4\frac{\theta}{2}+\frac{1}{N}\sin^4\frac{\theta}{2}\cr&&+
(\frac{N-1}{N}+\frac{2}{\sqrt{N}})\cos^2\frac{\theta}{2}\sin^2\frac{\theta}{2},
\end{eqnarray}
where $\theta\in[0,\pi/2]$. For the case $\theta$ = $\pi/2$ and $N$
= 3, we obtain the optimal fidelity of the $1$ $\rightarrow$ $3$
phase-covariant cloning ~\cite{JOB2005139}, therefore we construct a
$1$ $\rightarrow$ $N$ distributed orbital state quantum cloning
machine with only one step.

Our derivation for the effective reduced density operators of
different qubit is based on the quantum Zeno dynamics. To check the
validity of the result, we numerically simulate the effective model
in Eq. (11) and Eq. (12), and the full Hamiltonian model in Fig. 3
(a). We also plot the evolutions for reduced density operators of
the effective model and full model with different experimental
parameters $\theta$, $M$, $N$ versus $gt$ respectively from Fig. 3
(b) to Fig. 3 (g). The result shows that the smaller the $\theta$,
the fidelity of all copies is higher and fluctuates less in the
system evolution, and the operation time is proportional to
$\sqrt{M}$ in Fig. 3 (b) and (c). The tendencies in Fig. 3 (d) and
(e) indicate that the more the number of atoms $M$ in each cavity,
the time needed to reach the optimal fidelity of all copies becomes
longer. The plots of Fig. 3 (f) and (g) demonstrate that the result
of the quantum cloning become bad as the number of copies $N$
increases. Fig. 4 shows the influences of different fluctuations on
the fidelity $F$ of the cloned qubit $2$ when $N$ $=$ $3$.
Particularly, $F$ versus the errors of the interaction time $t$ and
the fluctuations in $\theta$ are plotted in Fig. 4 (g), the result
shows that the optimal fidelity of cloned qubit $2$ is almost
unaffected even when the relative fluctuations in $t$ and $\theta$
are both about $10\%$. $10\%$ deviations in other different
parameters only reduces the fidelity by about $10^{-2}$. The reason
is that our quantum state cloning is dependent on the evolution of
the $W$ state. However, the fidelity is independent of the phase
factor $\delta$ which makes the quantum state cloning workable.
Therefore, the distributed quantum cloning based on quantum Zeno
dynamics is robust against fluctuations of different experiment
parameters.

\section{Experiment feasibility and conclusion}

Now we give a detailed discussion of the experimental feasibility
and this secular setup. The atomic configuration involved in our
scheme can be implemented with a cesium atom \cite{PRL2004233603}.
For $7s_{1/2}$ of the cesium atom, we have $r_g$ $\approx$ $2.6nm$
and $d$ $\approx$ $4.1M^{-1/2}\mu m$, where the
no-direct-interaction condition can be satisfied very well when $M$
$<$ $200$. Thus, all the atoms in the cavity have the nearly same
coupling strength and collectively interact with the cavity field
\cite{PRA2007012324}. Single cesium atom can be localized at a fixed
position in each cavity with a high precision for a long time
\cite{PRL2003133602}. The near-perfect cavity-fiber coupling with an
efficiency larger than $99.9$\% has been reported
\cite{PRL91043902}, and a technique of using a $1$ $\times$ $N$ star
fiber optic coupler as a distributed strain sensor in a white-light
interferometer is presented \cite{AO19984168}. The star coupler is
realized as in \cite{IEEE1989241} by using a planar arrangement of
two confocal arrays of radial waveguides performing with efficiency
approaching $100\%$ under ideal conditions, when the waveguides have
strong mutual couplings. Finally, $N$ resonant classical laser
fields are applied to each atom within an appropriate operation
time, which plays the role of the `` external fields " for the
quantum Zeno dynamics. In fact, we do not think of $g$ and $v$ as
free control parameters which can be made arbitrarily `` large ",
but fix the parameters $g$ and $v$ during the evolution of the
system, and take the condition that $g$ and $v$ are both infinitely
strong just as an approximation \cite{JPCS2009012017}, compared to
$\Omega$. Even when the condition $\Omega$ $\ll$ $g,v$ is not well
satisfied, i.e., the system does not evolve via quantum Zeno
dynamics, the fidelity of the entanglement also keeps very high. For
example, even when $\Omega$ $=0.1\sqrt{M}g=18.5$MHz and $v$
$=0.1\sqrt{M}g=18.5$MHz, the fidelity of the entanglement is
$87.37\%$.

In conclusion, we have proposed a scheme to implement the
distributed orbital state quantum cloning with atomic ensembles via
quantum Zeno dynamics. The quantum cloning can be achieved via
appropriate coupling of atomic ensemble qubits at different nodes.
Therefore, the present scheme provides a new way to construct an
atomic ensemble quantum network via quantum Zeno dynamics using the
single-mode integrated optical star coupler, which is very robust
against cavity decay and very promising to be realized with current
experiment technology.

\section{Acknowledgements} \label{sec:4}
L.T.S. and H.Z.W acknowledge support from the Major State Basic
Research Development Program of China under Grant No. 2012CB921601,
National Natural Science Foundation of China under Grant No.
10974028, the Doctoral Foundation of the Ministry of Education of
China under Grant No. 20093514110009, and the Natural Science
Foundation of Fujian Province under Grant No. 2009J06002. Z.B.Y is
supported by the National Basic Research Program of China under
Grants No. 2011CB921200 and No. 2011CBA00200, and the China
Postdoctoral Science Foundation under Grant No. 20110490828.

\begin{figure}
\centering
\includegraphics[width=0.8\columnwidth]{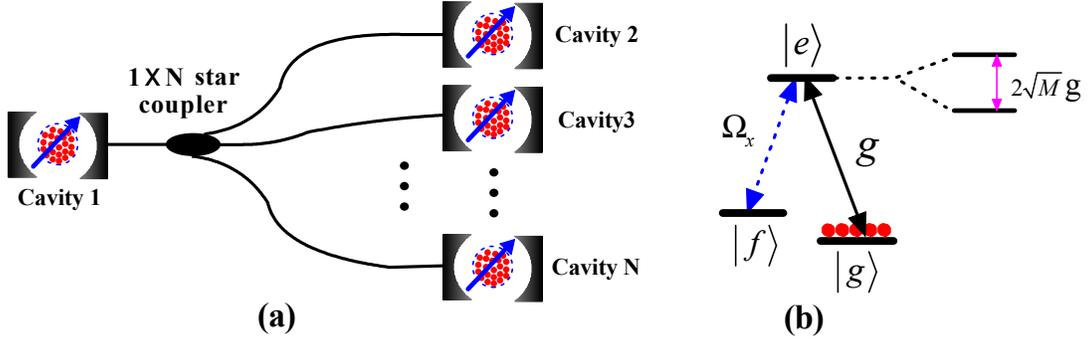} \caption{(Color online) (a) The experimental setup
  for implementing the distributed orbital state quantum cloning with atomic ensembles. \emph{M} identical three-level atoms are
  trapped in \emph{N} distant cavities respectively, and these cavities are connected by a 1 $\times$ $N$ single-mode integrated
  optical star coupler. Each blue solid in each cavity represents a classical
  laser on each atomic ensemble, which keeps invariant during the whole quantum Zeno
  dynamical evolution.
   (b) Configuration of the equivalent atomic ensemble level
  structure and relevant transitions. The states $|g\rangle$ and
  $|f\rangle$ correspond to two ground state of the atomic ensemble,
  and $|e\rangle$ is the excited state. $2\sqrt{M}g$ is the energy
  space between the dressed states of the excited state $|e\rangle$
  when the number of the atoms in the corresponding atomic ensemble is $M$.}
\end{figure}

\begin{figure}
\centering \subfigure[]{ \label{Fig.sub.a}
\includegraphics[width=0.5\columnwidth]{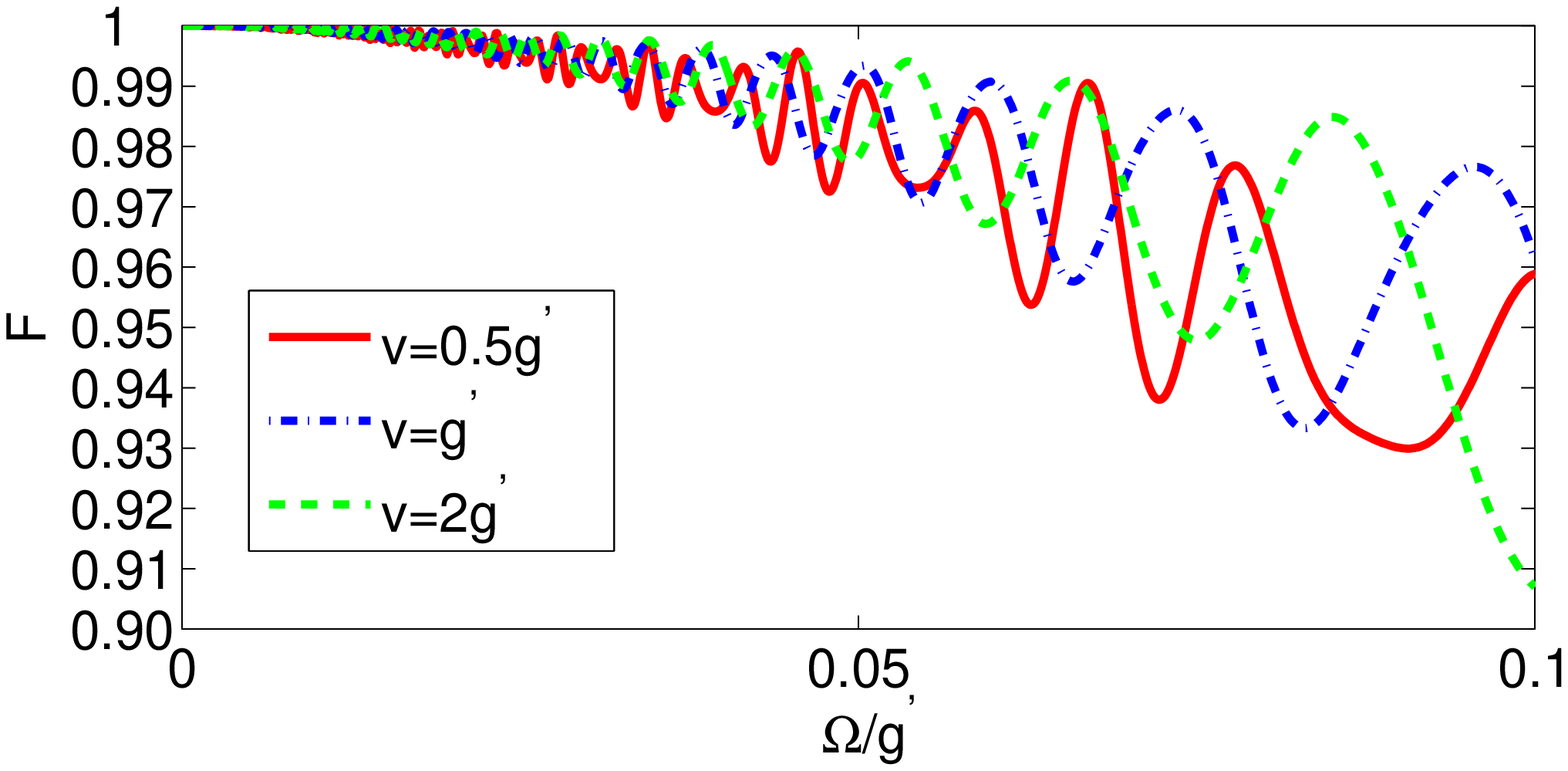}}
\subfigure[]{ \label{Fig.sub.b}
\includegraphics[width=0.5\columnwidth]{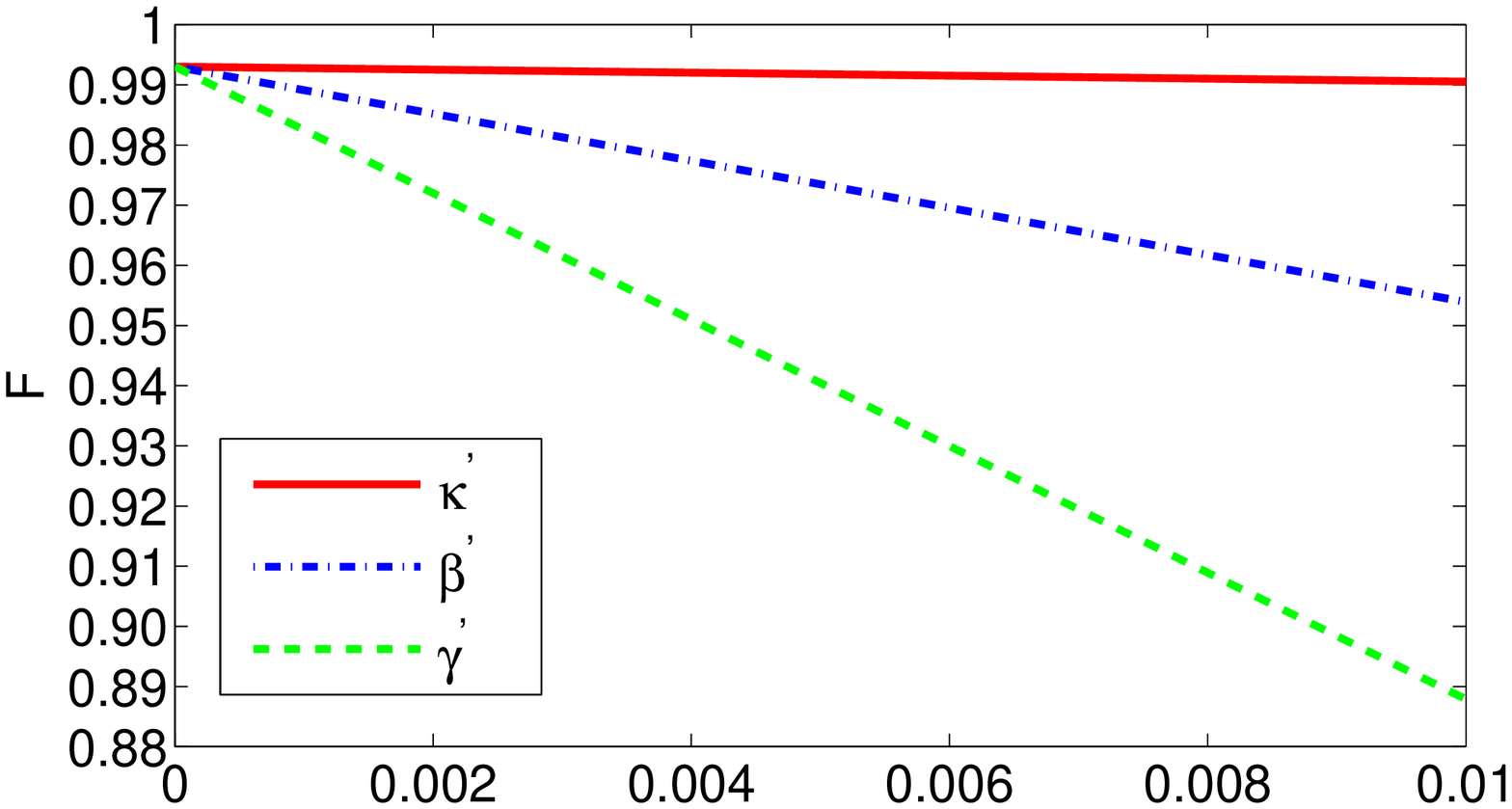}} \caption{(Color online) The fidelity
 $F$ of the entanglement between the atomic ensembles versus
 different parameters when $N$ = 3. (a) $F$ versus different ratios $\Omega/g'$ when $v=$ $0.5g'$ (the red solid line), $g'$ (the blue dot dash
 line) and $2g'$ (the green dash line), where $g'$ = $\sqrt{M}g$. (b) $F$ versus $\kappa'$ = $\kappa/(\sqrt{M}g)$ (the red solid line),
 $\beta'$ = $\beta/(\sqrt{M}g)$(the blue dot dash line), and $\gamma'$ = $\gamma/(\sqrt{M}g)$ (the green dash line).}
\end{figure}

\begin{figure}
\centering \subfigure[]{ \label{Fig.sub.a}
\includegraphics[width=0.35\columnwidth]{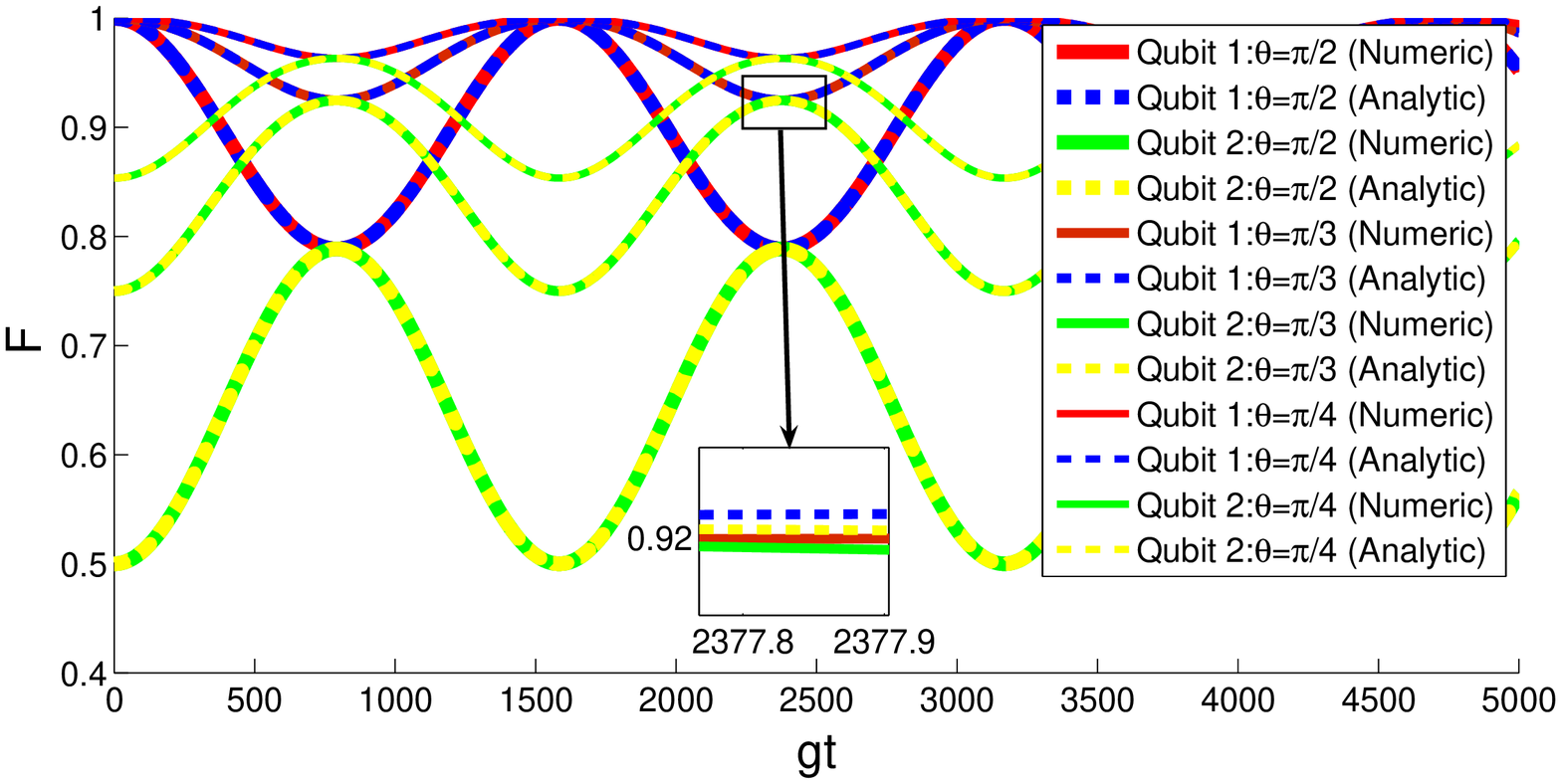}}\hspace{0.7in}
\centering \subfigure[]{ \label{Fig.sub.b}
\includegraphics[width=0.35\columnwidth]{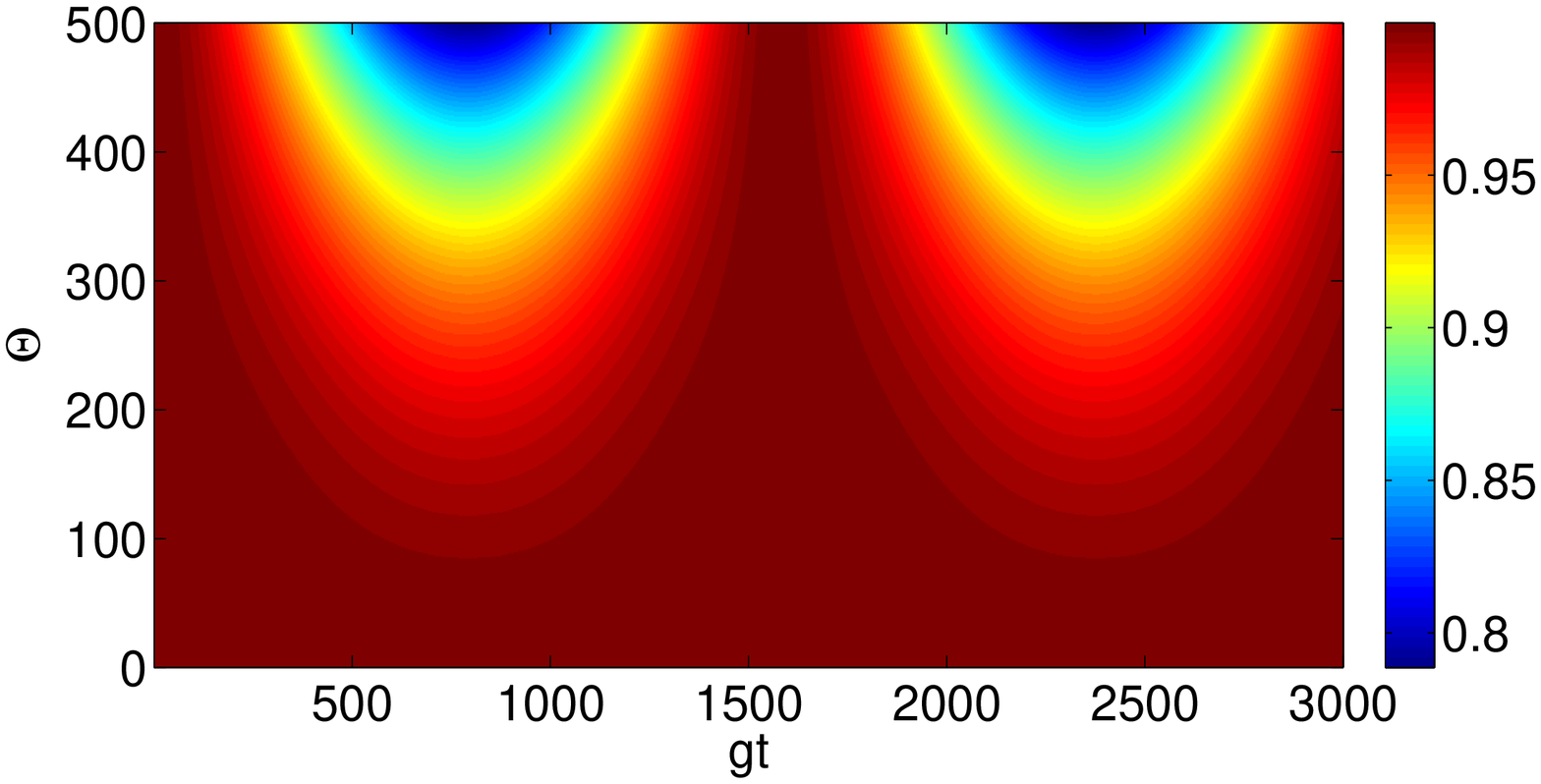}}
\centering \subfigure[]{ \label{Fig.sub.c}
\includegraphics[width=0.35\columnwidth]{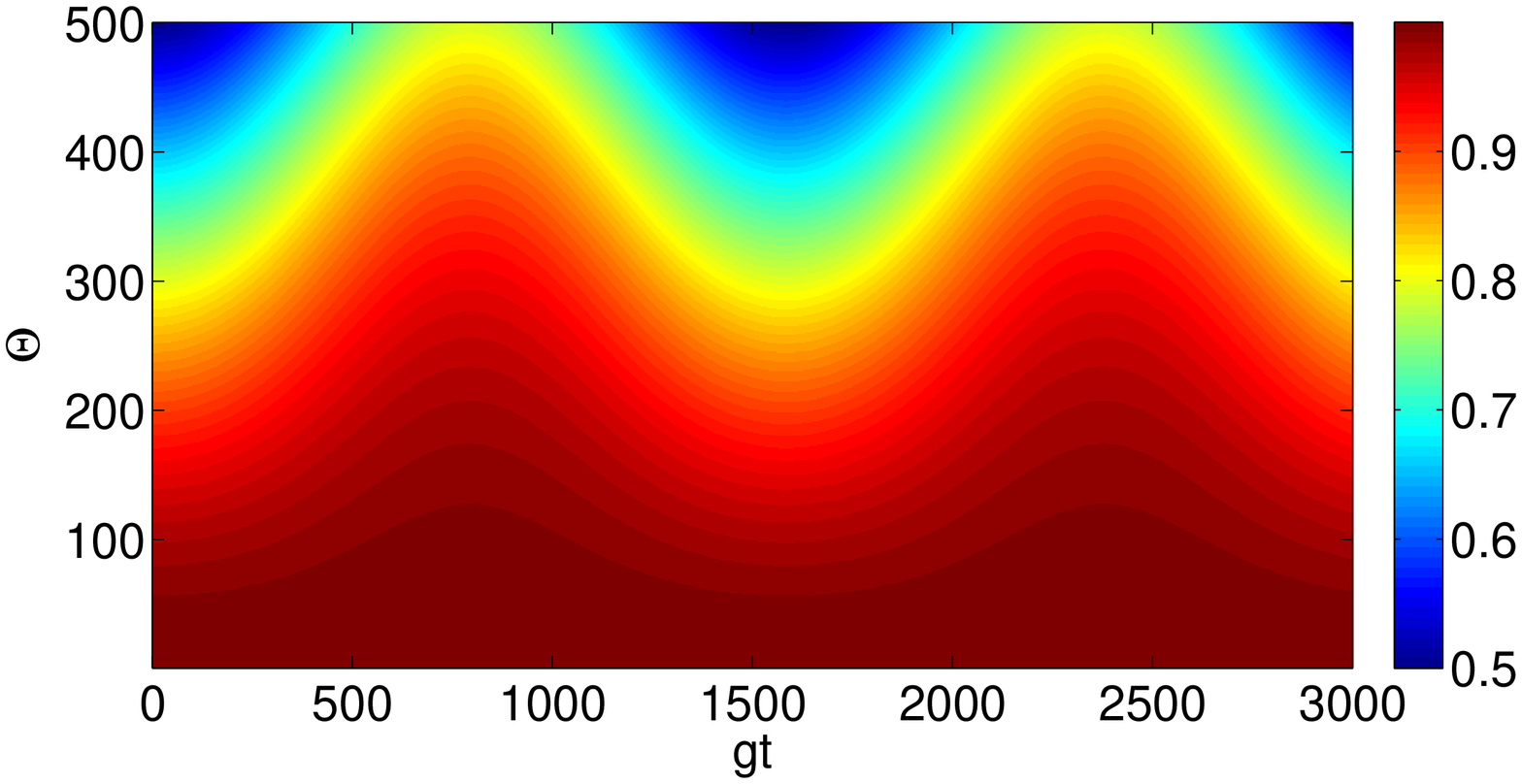}}\hspace{0.7in}
\centering \subfigure[]{ \label{Fig.sub.d}
\includegraphics[width=0.35\columnwidth]{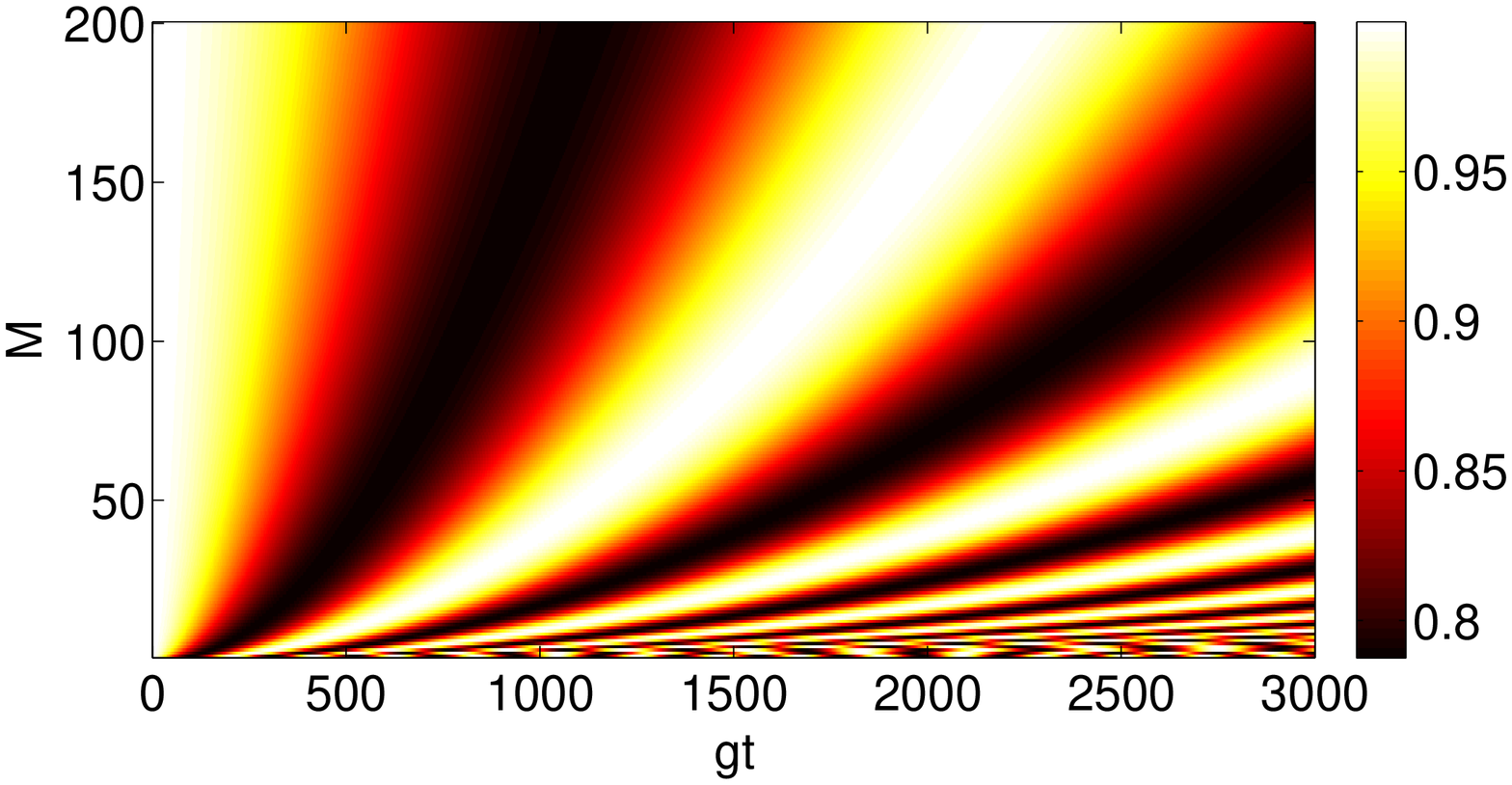}}
\centering \subfigure[]{ \label{Fig.sub.e}
\includegraphics[width=0.35\columnwidth]{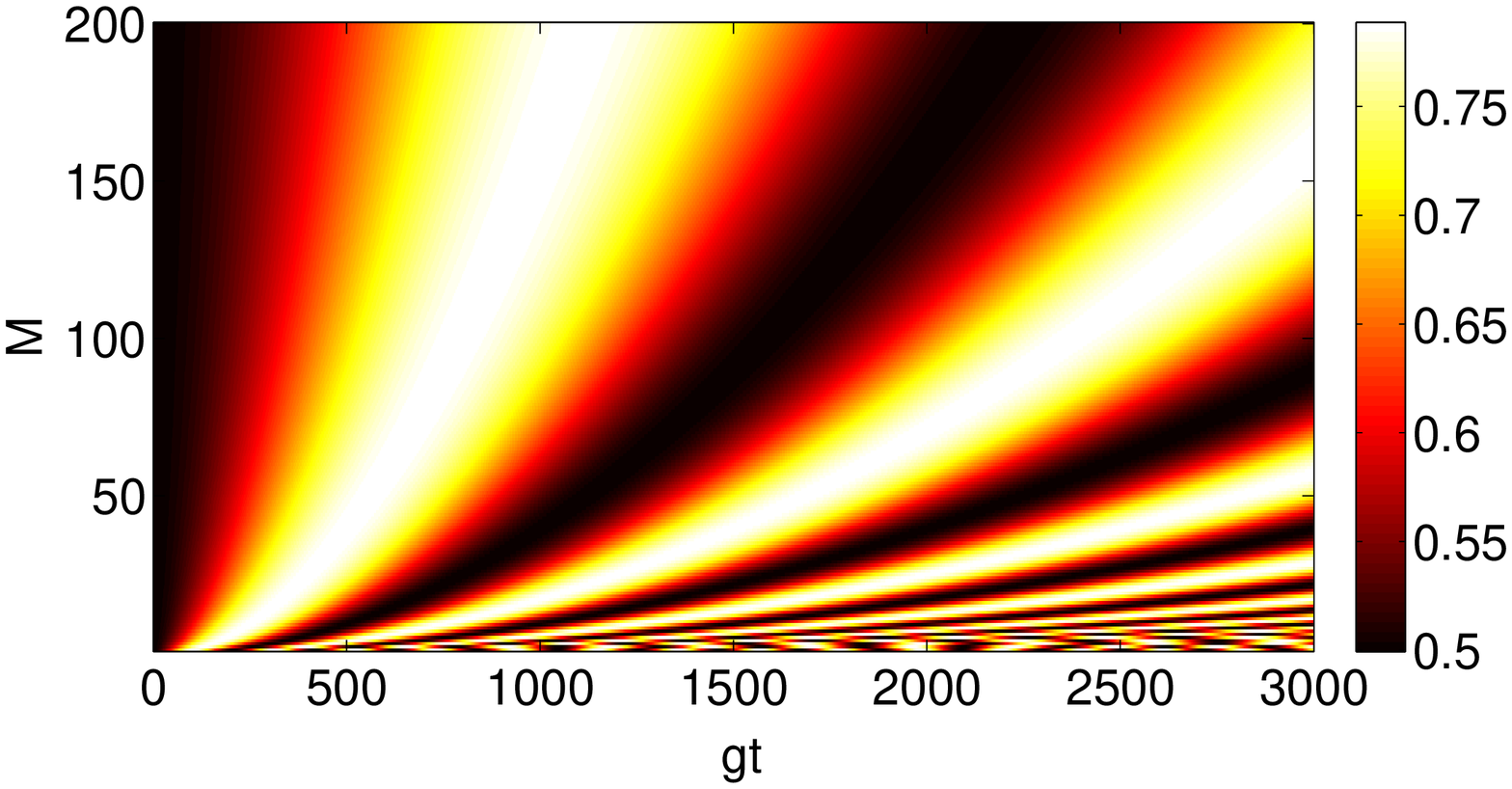}}\hspace{0.7in}
\centering \subfigure[]{ \label{Fig.sub.f}
\includegraphics[width=0.35\columnwidth]{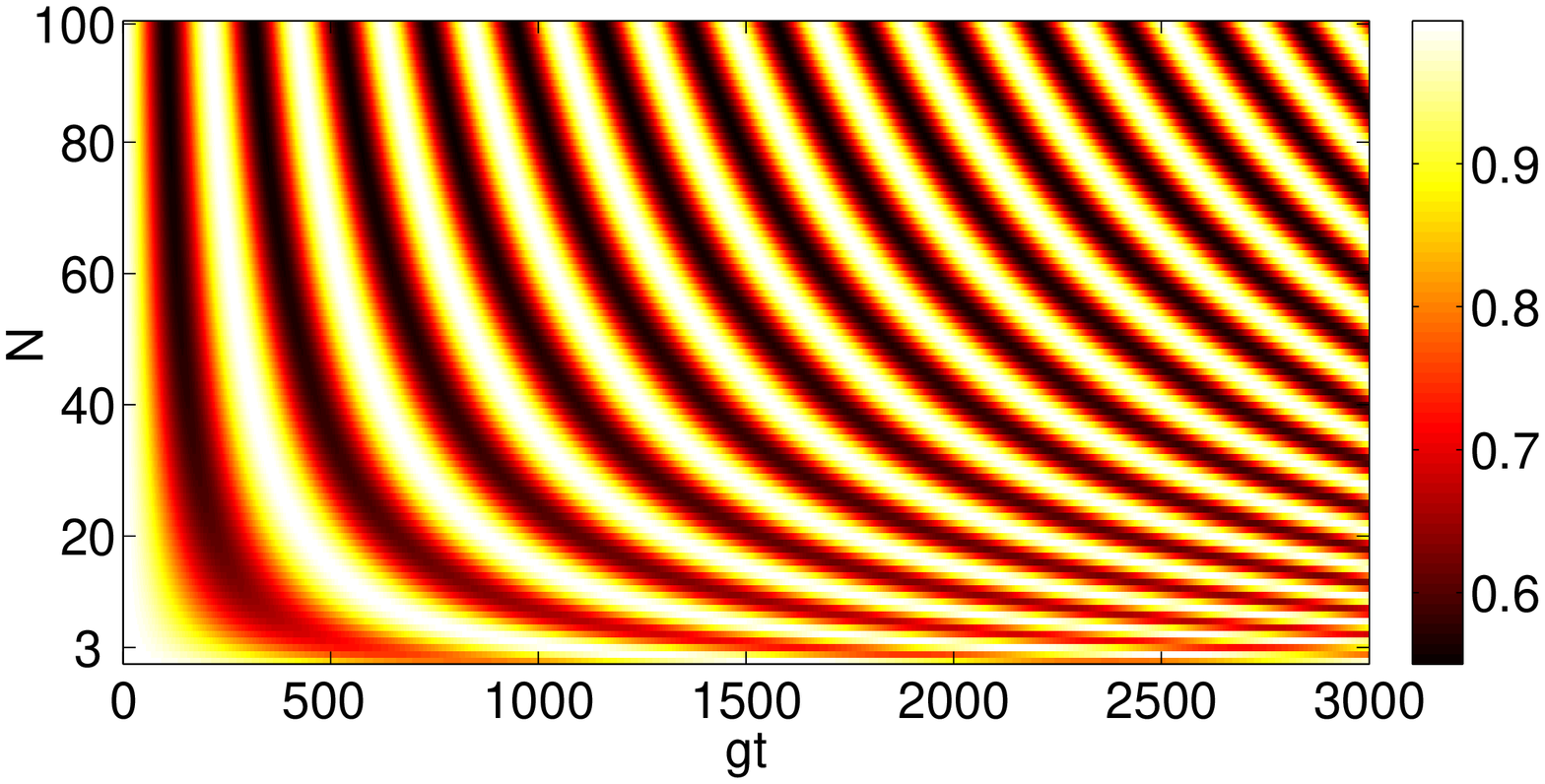}}
\centering \subfigure[]{ \label{Fig.sub.g}
\includegraphics[width=0.35\columnwidth]{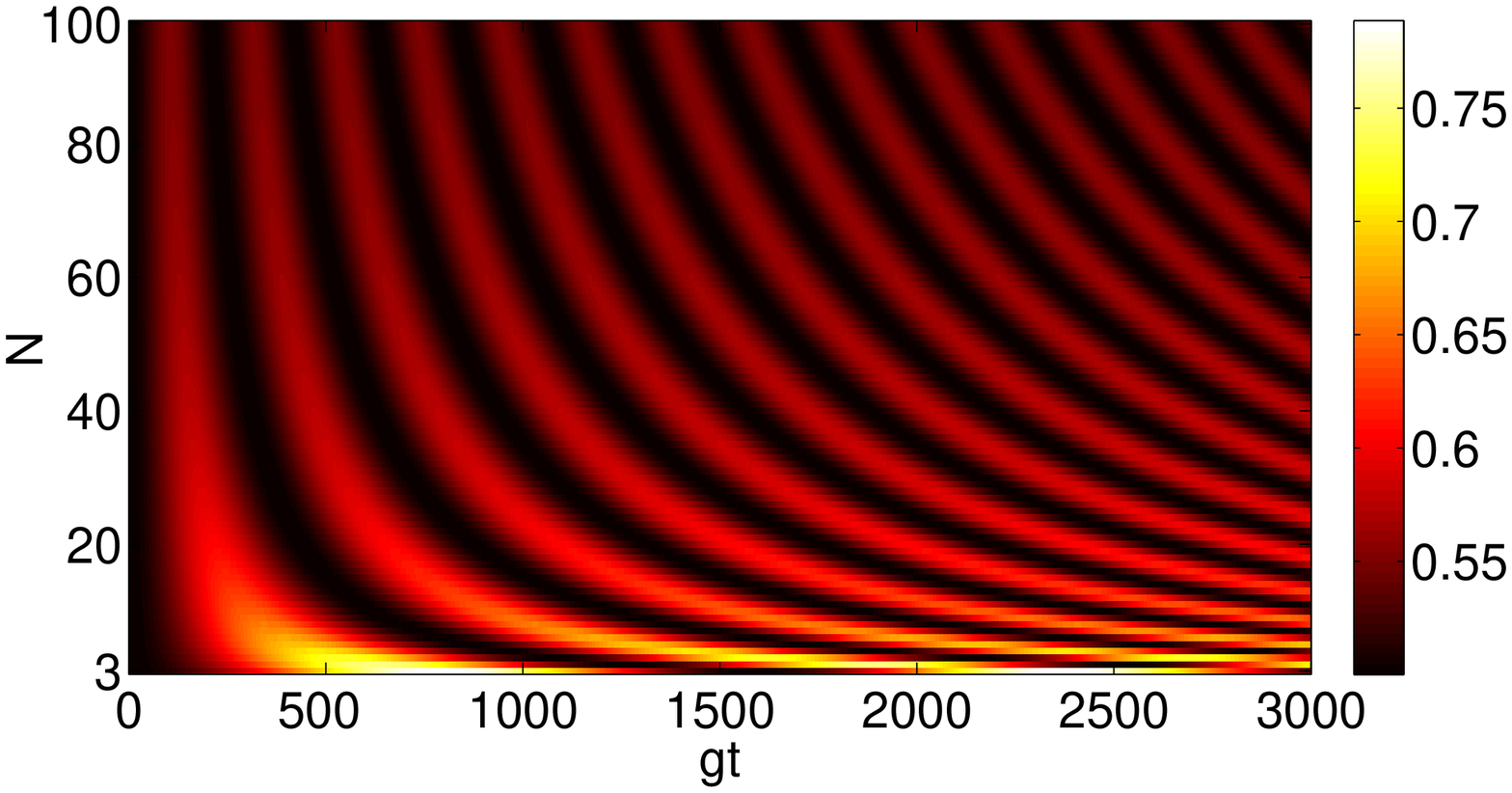}}
\centering \caption{(Color online) The effective model is plotted
with dash lines and the full dynamics is plotted with solid lines.
(a) The fidelity $F$ of different cloned qubits versus the
dimensionless parameter $gt$ with $N$ $=$ $3$ and $M=100$. (b) The
fidelity $F$ of qubit 1 versus parameters $\Theta$ ($\Theta$ $=$
$1000\theta/\pi$) and $gt$. (c) The fidelity $F$ of qubit 2 versus
parameters $\Theta$ ($\Theta$ $=$ $1000\theta/\pi$) and $gt$. (d)
The fidelity $F$ of qubit 1 versus parameters $M$ and $gt$ with
$\theta$ $=$ $\pi/2$ and $N$ $=$ $3$. (e) The fidelity $F$ of qubit
2 versus parameters $M$ and $gt$ with $\theta$ $=$ $\pi/2$ and $N$
$=$ $3$. (f) The fidelity $F$ of qubit 1 versus parameters $N$ and
$gt$ with $\theta$ $=$ $\pi/2$ and $M$ $=$ $100$. (g) The fidelity
$F$ of qubit 2 versus parameters $N$ and $gt$ with $\theta$ $=$
$\pi/2$ and $M$ $=$ $100$.}
\end{figure}

\begin{figure}[htbp]
\centering 
\subfigure[]{ \label{Fig.sub.a}
\includegraphics[width=0.25\columnwidth]{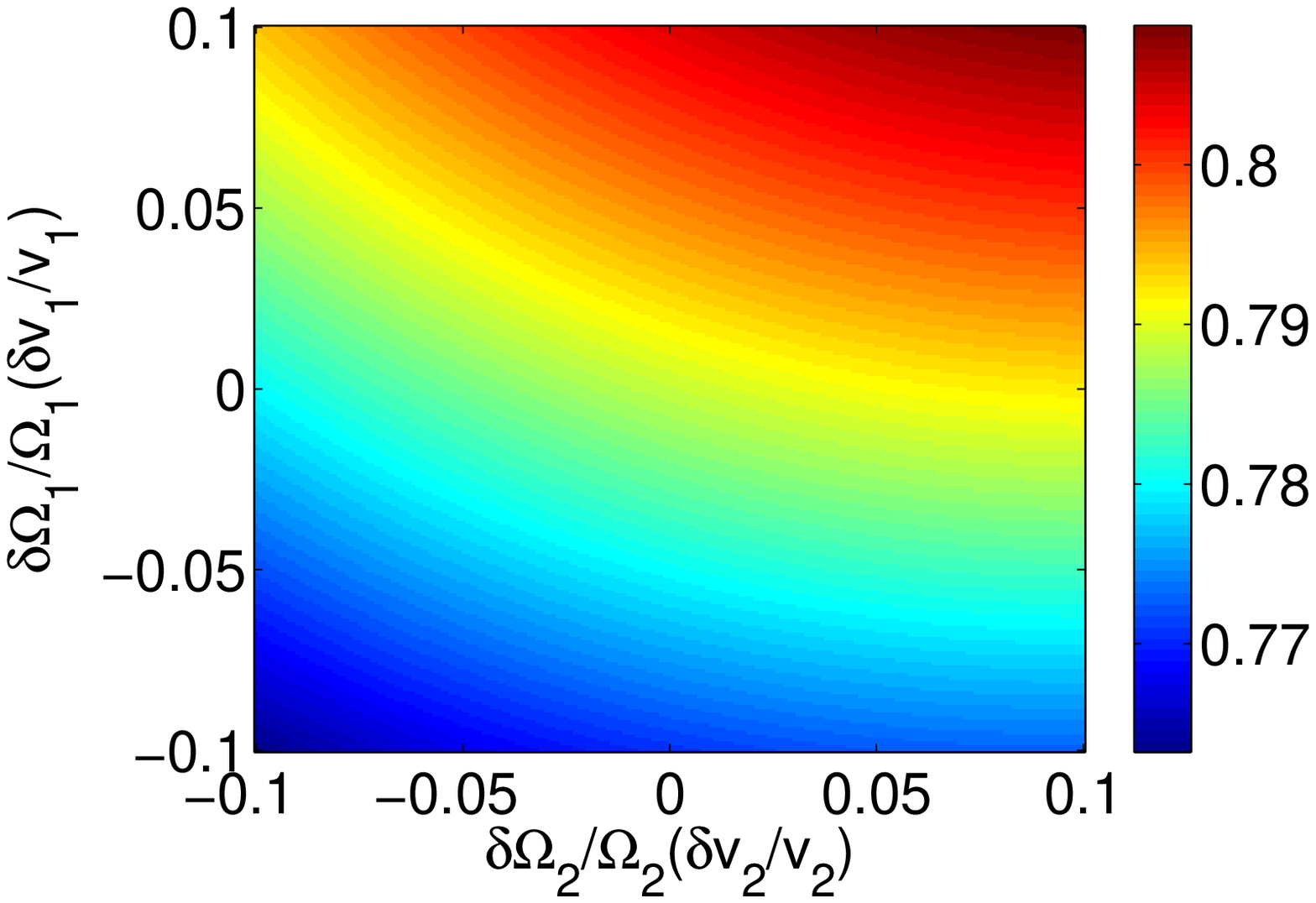}}\hspace{1.5in}
\subfigure[]{ \label{Fig.sub.b}
\includegraphics[width=0.25\columnwidth]{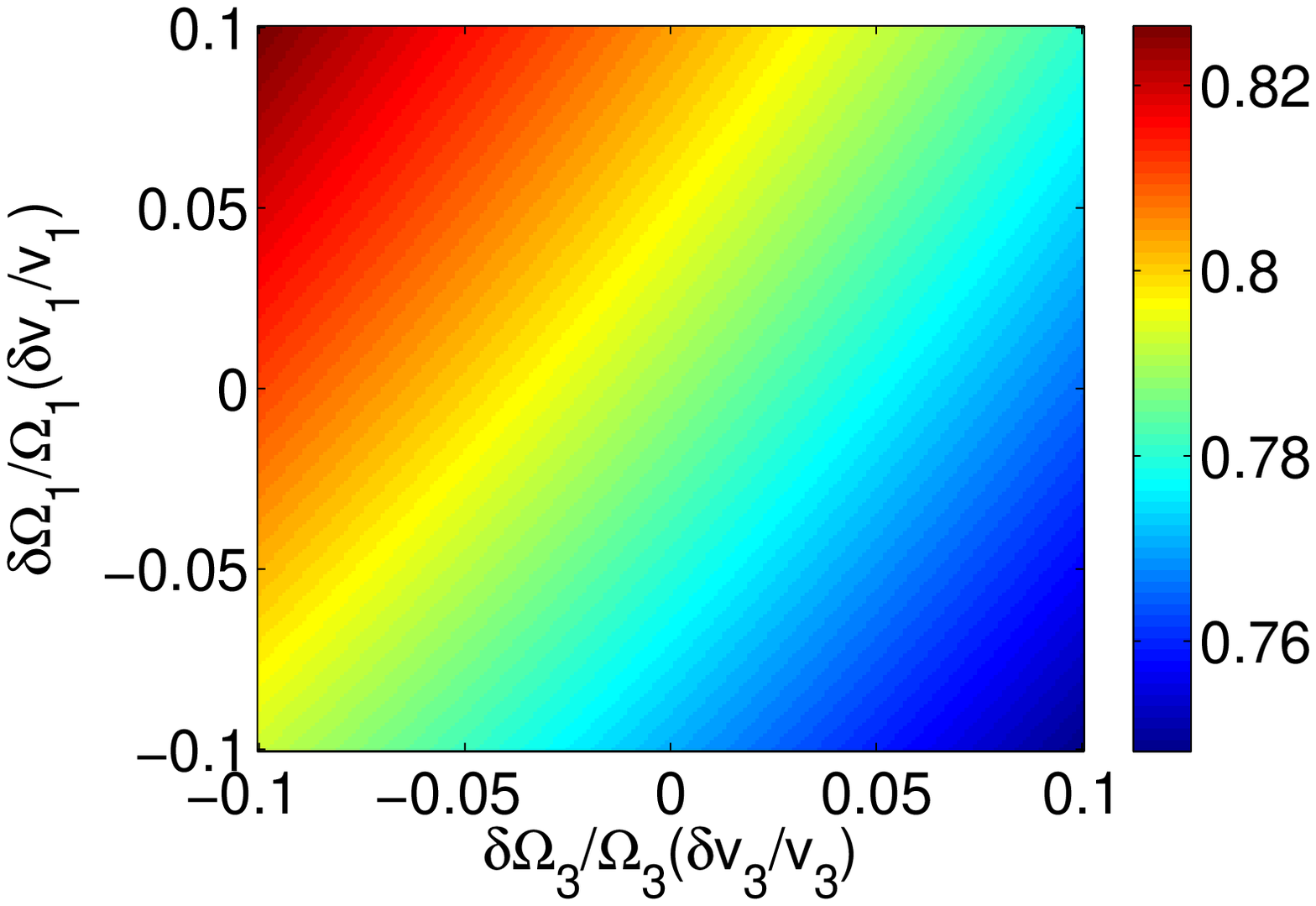}}\hspace{1.5in}
\subfigure[]{ \label{Fig.sub.c}
\includegraphics[width=0.25\columnwidth]{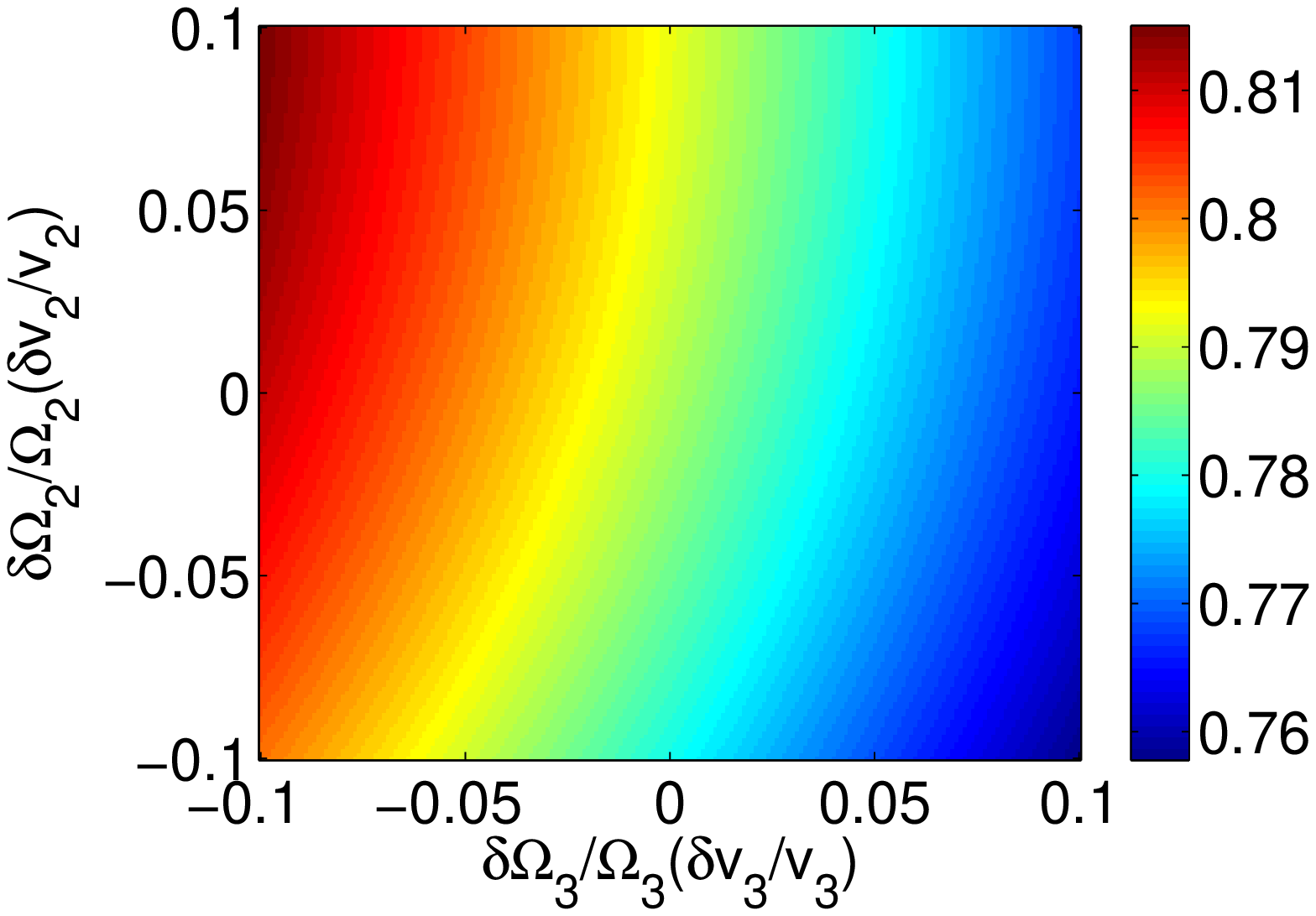}}\hspace{1.5in}
\subfigure[]{ \label{Fig.sub.d}
\includegraphics[width=0.25\columnwidth]{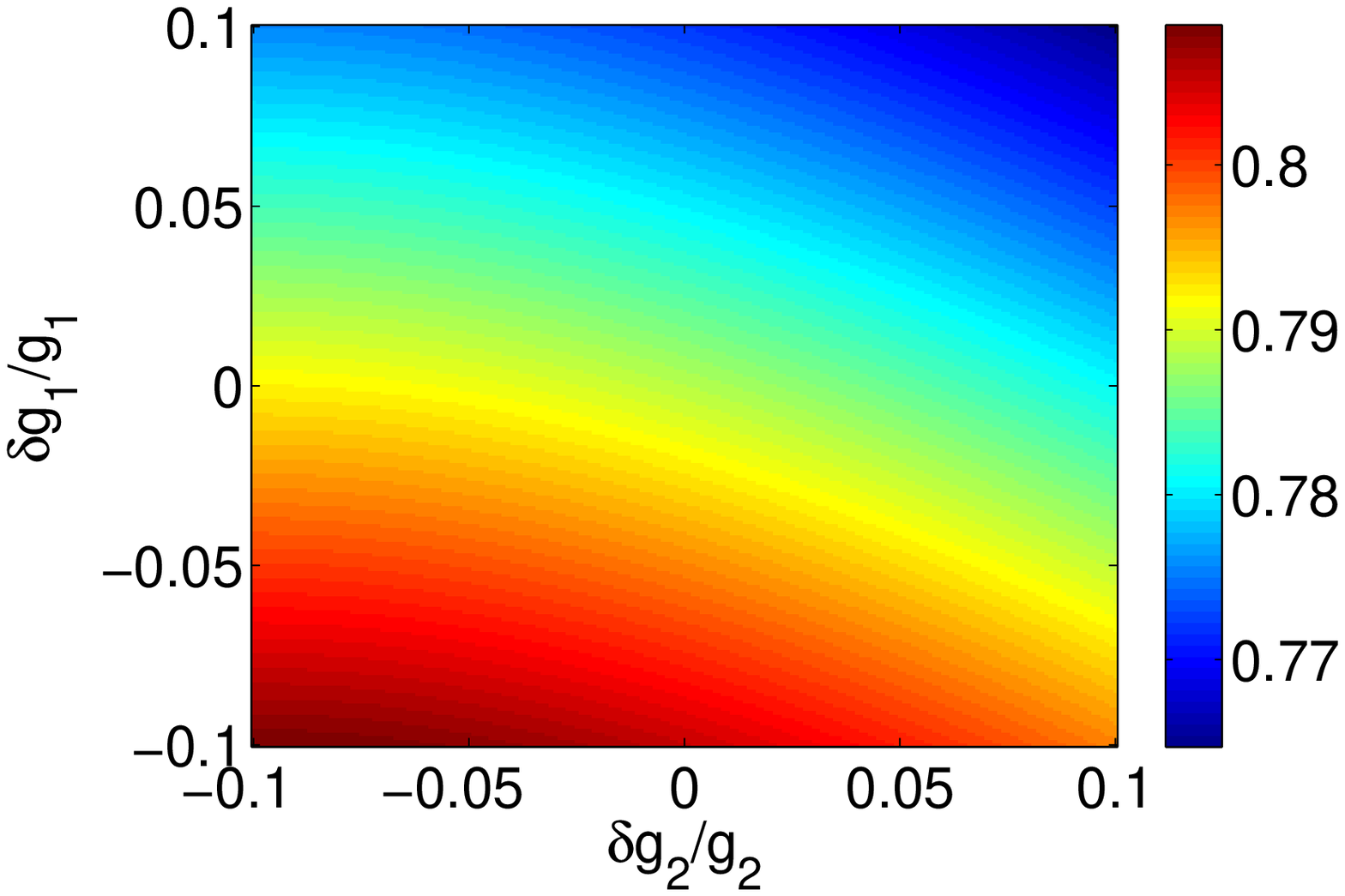}}\hspace{1.5in}
\subfigure[]{ \label{Fig.sub.e}
\includegraphics[width=0.25\columnwidth]{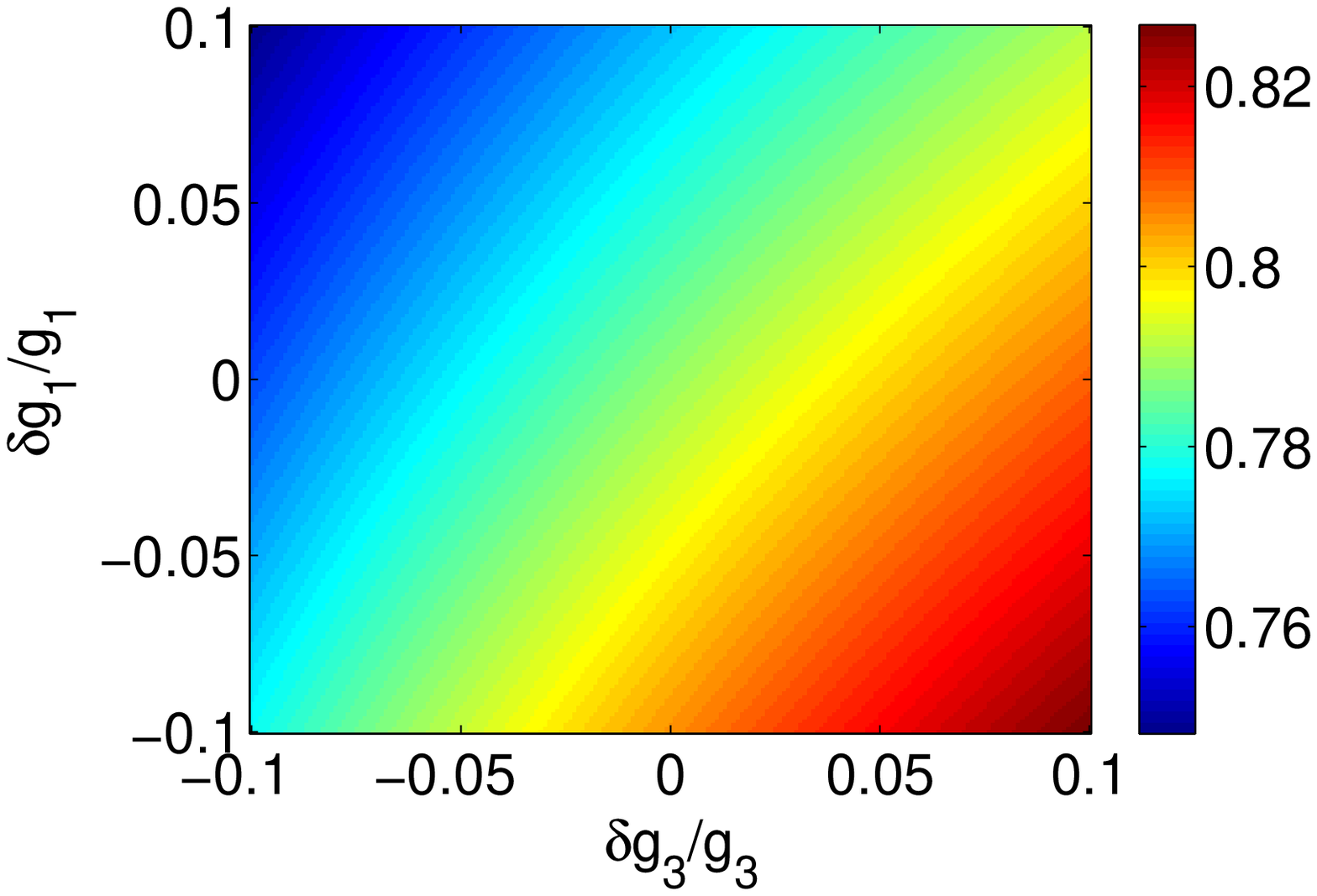}}\hspace{1.5in}
\subfigure[]{ \label{Fig.sub.f}
\includegraphics[width=0.25\columnwidth]{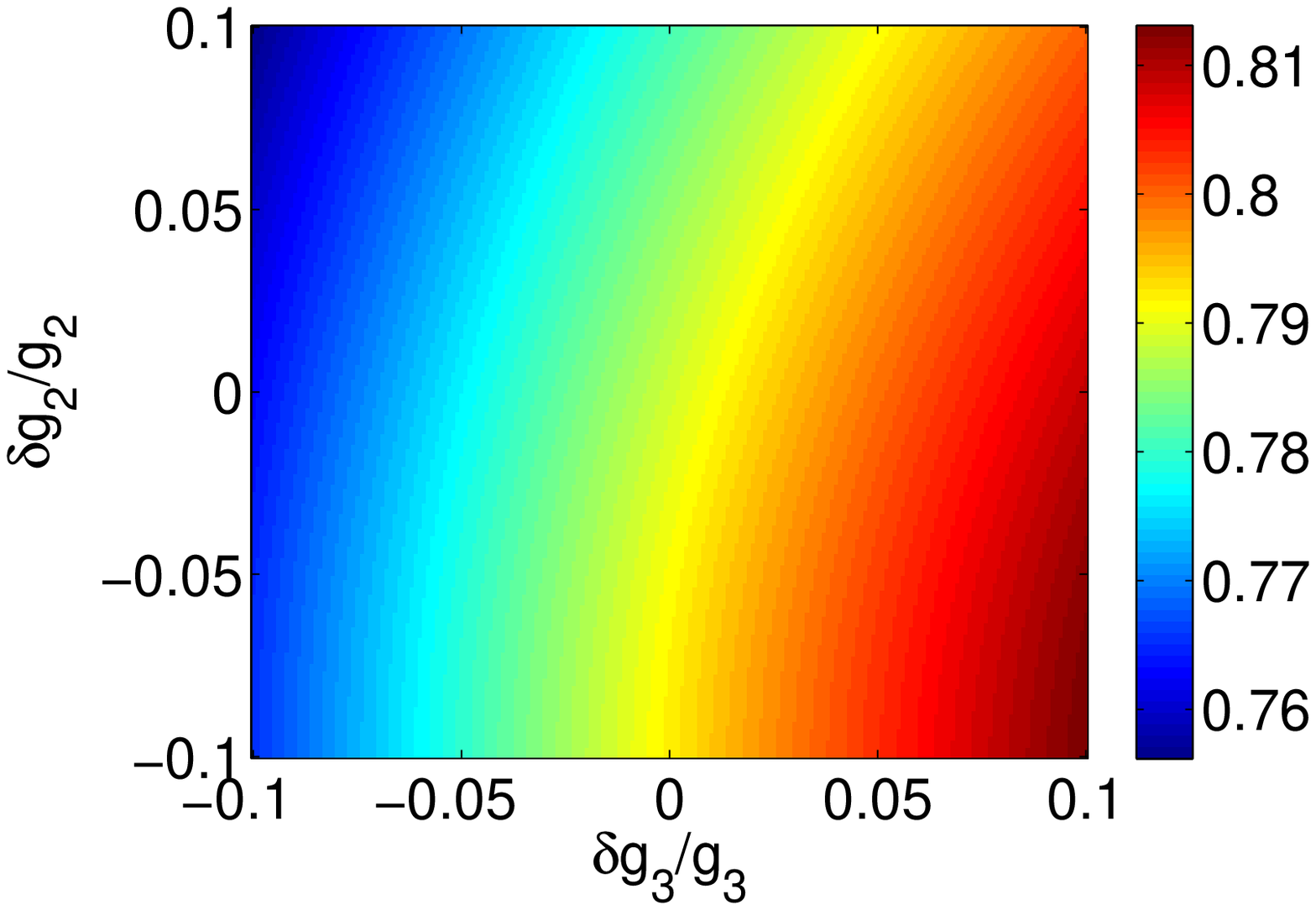}}\hspace{1.5in}
\subfigure[]{ \label{Fig.sub.g}
\includegraphics[width=0.25\columnwidth]{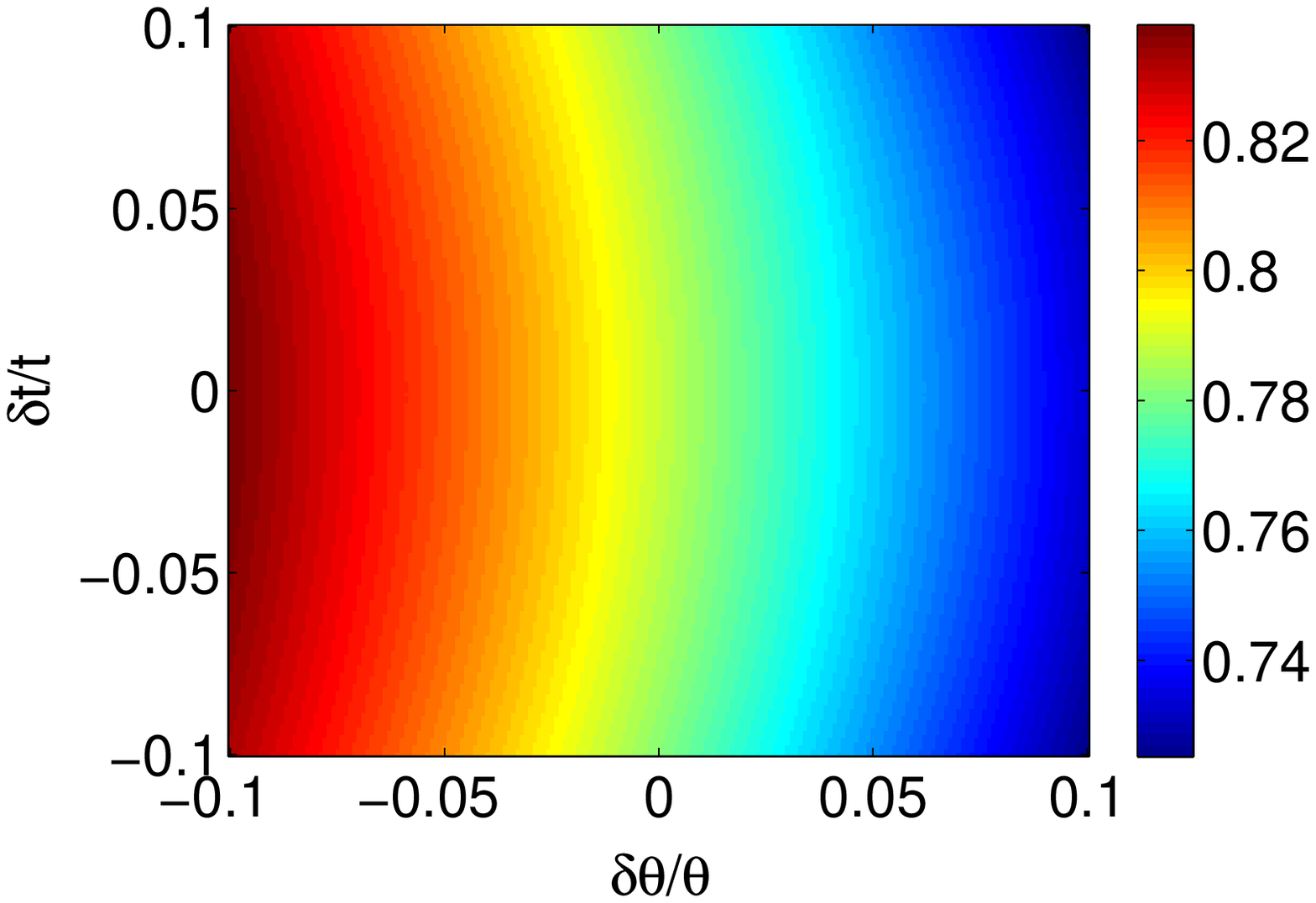}}\hspace{1.5in}
 \caption{(Color online) The
fidelity $F$ of the cloned qubit 2 versus various fluctuations of
the experiment parameters. All subfigures are plotted with the
parameter $N$ $=$ $3$, $\Omega$ $=$ $0.05\sqrt{M}g$, $v$ $=$ $0.5g$
and $\theta$ $=$ $\pi/2$, therefore the optimal fidelity is $0.788$.
$g_1$ $=$ $g_2$ $=$ $g_3$ $=$ $\sqrt{M}g$. (a) $F$ vs
$\frac{\delta\Omega_1(\delta v_1)}{\Omega_1(v_1)}$ and
$\frac{\delta\Omega_2(\delta v_2)}{\Omega_2(v_2)}$; (b) $F$ vs
$\frac{\delta\Omega_1(\delta v_1)}{\Omega_1(v_1)}$ and
$\frac{\delta\Omega_3(\delta v_3)}{\Omega_3(v_3)}$; (c) $F$ vs
$\frac{\delta\Omega_2(\delta v_2)}{\Omega_2(v_2)}$ and
$\frac{\delta\Omega_3(\delta v_3)}{\Omega_3(v_3)}$; (d) $F$ vs
$\frac{\delta g_1}{g_1}$ and $\frac{\delta g_2}{g_2}$; (e) $F$ vs
$\frac{\delta g_1}{g_1}$ and $\frac{\delta g_3}{g_3}$; (f) $F$ vs
$\frac{\delta g_2}{g_2}$ and $\frac{\delta g_3}{g_3}$; (g) $F$ vs
$\frac{\delta t}{t}$ and $\frac{\delta \theta}{\theta}$;}
\end{figure}

\newpage \nonumber


\begin{thebibliography}{90}
\bibitem{PLA2000275} P.~Facchi, V.~Gorini, G.~Marmo, S.~Pascazio, and E.~C.~G.~Sudarshan, Phys. Lett. A \textbf{275}, 12 (2000).
\bibitem{JMP1997756} B.~Misra and E.~C.~G.~Sudarshanand, J. Math. Phys. \textbf{18}, 756 (1977).

\bibitem{PRA199041} W.~M.Itano, D.~J.~Heinzen, J.~J.~Bollinger, and D.~J.~Wineland, Phys. Rev. A \textbf{41}, 2295 (1990).
\bibitem{PRL200289} P.~Facchi and S.~Pascazio, Phys. Rev. Lett. \textbf{89}, 080401 (2002).
\bibitem{JPCS2009012017} P.~Facchi, G.~Marmo, and S.~Pascazio, J. Phys: Conf. Ser. \textbf{196}, 012017 (2009).%
\bibitem{PRA2008062332} Y.~P.~Huang and M.~G.~Moore, Phys. Rev. A \textbf{77}, 062332 (2008).
\bibitem{PRA2003022320} H.~Azuma, Phys. Rev. A \textbf{68}, 022320 (2003).
\bibitem{PRA2007052339} C.~R.~Myers and A.~Gilchrist, Phys. Rev. A \textbf{75}, 052339 (2007).

\bibitem{EPL201050003}  X.~Q.~Shao, L.~Chen, S.~Zhang, Y.~F.~Zhao, and K.~H.~Yeon, Eur. Phys. Lett. \textbf{90}, 50003 (2010).
\bibitem{PRA2009062323} X.~Q.~Shao, H.~F.~Wang, L.~Chen, S.~Zhang, Y.~F.~Zhao, and K.~H.~Yeon, Phys. Rev. A \textbf{80}, 062323 (2009).
\bibitem{PRA2011022322} W.~A.~Li and G.~Y.~Huang, Phys. Rev. A \textbf{83}, 022322 (2011).
\bibitem{PRT198821} R.~J.~Cook, Phys. Scr. T \textbf{21}, 49 (1988).    
\bibitem{PRL2008180402} J.~Bernu, S.~Del$\acute{e}$glise, C.~Sayrin, S.~Kuhr, I.~Dotsenko, M.~Brune, J.~M.~Raimond, and S.~Haroche, Phys. Rev.Lett. \textbf{101}, 180402 (2008).
\bibitem{PRL2006260402} E.~W.~Streed, J.~Mun, M.~Boyd, G.~K.~Campbell, P.~Medley, W.~Ketterle, and D.~E.~Pritchard, Phys. Rev. Lett. \textbf{97}, 260402 (2006).

\bibitem{PRL19975242} T.~Pellizzari, Phys. Rev. Lett. \textbf{79}, 5242 (1997).
\bibitem{APL2009154101} S.~B.~Zheng, Appl. Phys. Lett. \textbf{90}, 154101 (2009).
\bibitem{PRA2007012324} Z.~Q.~Yin and F.~L.~Li, Phys. Rev. A 75, 012324 (2007).
\bibitem{PRL2006010503} A.~Serafini, S.~Mancini, and S.~Bose, Phys. Rev. Lett. \textbf{96}, 010503 (2006).
\bibitem{PRA2009012305} Z.~B.~Yang, H.~Z.~Wu, W.~J.~Su, and S.~B.~Zheng, Phys. Rev. A \textbf{80}, 012305 (2009).
\bibitem{PRA2008014303} S.~Y.~Ye, Z.~R.~Zhong, and S.~B.~Zheng, Phys. Rev. A \textbf{77}, 014303 (2008).
\bibitem{PRA2010042327} S.~B.~Zheng, C.~P.~Yang, and F.~Nori, Phys. Rev. A \textbf{82}, 042327 (2010).
\bibitem{PRA2007062320} P.~Peng and F.~L.~Li, Phys. Rev. A \textbf{75}, 062320 (2007).
\bibitem{EPL200760001}  J.~Song, Y.~Xia, H.~S.~Song, J.~L.~Guo, and J.~Nie, Eur. Phys. Lett. \textbf{80}, 60001 (2007).
\bibitem{PRL2004210501} D.~Bru{\ss}, G.~M.~D'Ariano, M.~Lewenstein, C.~Macchiavello,
A.~Sen(De), and U.~Sen, Phys. Rev. Lett. \textbf{93}, 210501 (2004).
\bibitem{Nature1982299} W.~K.~Wootters and W.~H.~Zurek, Nature (London) \textbf{299}, 802 (1982).%
\bibitem{PRA2000012302} D.~Bru${\ss}$, M.~Cinchetti, G.~M.~D'Ariano,
and C.~Macchiavello, Phys. Rev. A \textbf{62}, 012302 (2000).
\bibitem{PRA2001012304} H.~Fan, K.~Matsumoto, X.~B.~Wang, and
M.~Wadati, Phys. Rev. A \textbf{65}, 012304 (2001).
\bibitem{PRA2003042306} G.~M.~D'Ariano and C.~Macchiavello, Phys. Rev. A \textbf{67}, 042306
(2003).

\bibitem{PRA199858} R.~F.~Werner, Phys. Rev. A \textbf{58}, 1827 (1998).
\bibitem{PRA199654} V.~Bu$\breve{z}$ek and M.~Hillery, Phys. Rev. A \textbf{54}, 1844 (1996).%
\bibitem{PRL199779} N.~Gisin and S.~Massar, Phys. Rev. Lett. \textbf{79}, 2153 (1997).
\bibitem{JOB2005139} S.~B.~Zheng, J. Opt. B: Quantum Semiclassical Opt. \textbf{7}, 139 (2005).
\bibitem{Nature2011210} J.~Volz, R.~Gehr, G.~Dubois, J.~Est$\grave{e}$ve, and J.~Reichel, Nature \textbf{475}, 210 (2011).
\bibitem{IEEE2004900} K.~J.~Gordon, V.~Fernandez, P.~D.~Townsend, and G.~S.~Buller, IEEE J. Quantum Electron. \textbf{40}, 900 (2004).
\bibitem{PLA2005278}  A.~T.~Rezakhani, S.~Siadatnejad, and A.~H.~Ghaderi, Phys. Lett. A \textbf{336}, 278 (2005).
\bibitem{Science2002296} A.~Lamas-Linares, C.~Simon, J.~C.Howell, and D.~Bouwmeester, Science \textbf{296}, 712 (2002). 
\bibitem{PRL2002187901} H.~K.Cummins, C.~Jones, A.~Furze, N.~F.Soffe, M.~Mosca, J.~M.~Peach, and J.~A.~Jones, Phys. Rev. Lett. \textbf{88}, 187901 (2002).
\bibitem{PRL2005040505} J.~F.~Du, T.~Durt, P.~Zou, H.~Li, L.~C.~Kwek, C.~H.~Lai, C.~H.~Oh, and A.~Ekert, Phys. Rev. Lett. \textbf{94}, 040505 (2005).
\bibitem{PRL2011180404} H.~W.~Chen, D.~W.~Lu, B.~Chong, G.~Qin, X.~Y.~Zhou, X.~H.~Peng, and J.~F.~Du, Phys. Rev. Lett. \textbf{106}, 180404 (2011).
\bibitem{PRL2007170503} M.~Sabuncu, U.L.~Andersen, and G.~Leuchs, Phys. Rev. Lett. \textbf{98}, 170503 (2007).
\bibitem{PRL2010073602} E.~Nagali, D.~Giovannini, L.~Marrucci, S.~Slussarenko, E.~Santamato, and F.~Sciarrino, Phys. Rev. Lett. \textbf{105}, 073602 (2010).
\bibitem{Nature200653} J.~P.~Cleuziou \emph{et al.}, Nature \textbf{1}, 53 (2006).
\bibitem{Nature2010249} P.~Neumann, \emph{et al.}, Nature \textbf{6}, 249 (2010).
\bibitem{PRL19982598} D.~Bruss, A.~Ekert and C.~Macchiavello Phys. Rev. Lett. \textbf{81}, 2598 (1988).
\bibitem{PRL1991661} A.~K.~Ekert, Phys. Rev. Lett. \textbf{67}, 661 (1991).
\bibitem{PRA2001042308} D.~Bru{\ss}, J.~Calsamiglia and
N.~L\"{u}tkenhaus Phys. Rev. A \textbf{63}, 042308 (2001).

\bibitem{PRL2007120505} L.~Bart\r{u}\u{s}kov\'{a}, M.~Du\u{s}ek, A.~\u{C}ernoch, J.~Soubusta, and J.~Fiur\'{a}\u{s}ek, Phys. Rev. Lett. \textbf{99}, 120505 (2007).

\bibitem{PRA2006032329} S.~B.~Zheng and G.~C.~Guo, Phys. Rev. A \textbf{73}, 032329 (2006).


\bibitem{APL1982549} T.~Findakly and B.~Chen, Appl. Phys. Lett. \textbf{40}, 549 (1982).
\bibitem{JPNJAP19975136} F.~H, L.~E.Herbert, and T.~Kunio, Jpn. J. Appl.
Phys. \textbf{36}, 5136 (1997).

\bibitem{PRL2004233603} A.~Boca, R.~Miller, K.~M.~Birnbaum,
A.~D.~Boozer, J.~McKeever, and H.~J.~Kimble, Phys. Rev. Lett. \textbf{93}, 233603 (2004).
\bibitem{PRL2003133602} J.~McKeever, J.~R.~Buck, A.~D.~Boozer, A.~Kuzmich, H.~C.~Nagerl, D.~M.~Stamper-Kurn, and H.~J.~Kimble, Phys. Rev. Lett. \textbf{90}, 133602 (2003).
\bibitem{PRL91043902}  S.~M.~Spillane, T.~J.~Kippenberg, O.~J.~Painter, and
K.~J.~Vahala, Phys. Rev. Lett. \textbf{91}, 043902 (2003).

\bibitem{AO19984168} L.~B.~Yuan and L.~M.~Zhou, Appl. Opt. \textbf{90}, 4168 (1998).
\bibitem{IEEE1989241} C.~Dragone, C.~H.~Henry, I.~P.~Kaminow, and
R.~C.~Kistler, IEEE Photon. Technol. Lett. \textbf{1}, 241 (1989).
%
\end{thebibliography}
\end{document}